\documentclass[usenatbib,usegraphicx]{mn2e}
\usepackage{amsmath}
\usepackage{amsfonts}
\usepackage{amssymb}
\usepackage{comment}
\usepackage{color}
\definecolor{AliceBlue}{rgb}{0.94,0.97,1.00}
\definecolor{AntiqueWhite1}{rgb}{1.00,0.94,0.86}
\definecolor{AntiqueWhite2}{rgb}{0.93,0.87,0.80}
\definecolor{AntiqueWhite3}{rgb}{0.80,0.75,0.69}
\definecolor{AntiqueWhite4}{rgb}{0.55,0.51,0.47}
\definecolor{AntiqueWhite}{rgb}{0.98,0.92,0.84}
\definecolor{BlanchedAlmond}{rgb}{1.00,0.92,0.80}
\definecolor{BlueViolet}{rgb}{0.54,0.17,0.89}
\definecolor{CadetBlue1}{rgb}{0.60,0.96,1.00}
\definecolor{CadetBlue2}{rgb}{0.56,0.90,0.93}
\definecolor{CadetBlue3}{rgb}{0.48,0.77,0.80}
\definecolor{CadetBlue4}{rgb}{0.33,0.53,0.55}
\definecolor{CadetBlue}{rgb}{0.37,0.62,0.63}
\definecolor{CornflowerBlue}{rgb}{0.39,0.58,0.93}
\definecolor{DarkBlue}{rgb}{0.00,0.00,0.55}
\definecolor{DarkCyan}{rgb}{0.00,0.55,0.55}
\definecolor{DarkGoldenrod1}{rgb}{1.00,0.73,0.06}
\definecolor{DarkGoldenrod2}{rgb}{0.93,0.68,0.05}
\definecolor{DarkGoldenrod3}{rgb}{0.80,0.58,0.05}
\definecolor{DarkGoldenrod4}{rgb}{0.55,0.40,0.03}
\definecolor{DarkGoldenrod}{rgb}{0.72,0.53,0.04}
\definecolor{DarkGray}{rgb}{0.66,0.66,0.66}
\definecolor{DarkGreen}{rgb}{0.00,0.39,0.00}
\definecolor{DarkGrey}{rgb}{0.66,0.66,0.66}
\definecolor{DarkKhaki}{rgb}{0.74,0.72,0.42}
\definecolor{DarkMagenta}{rgb}{0.55,0.00,0.55}
\definecolor{DarkOliveGreen1}{rgb}{0.79,1.00,0.44}
\definecolor{DarkOliveGreen2}{rgb}{0.74,0.93,0.41}
\definecolor{DarkOliveGreen3}{rgb}{0.64,0.80,0.35}
\definecolor{DarkOliveGreen4}{rgb}{0.43,0.55,0.24}
\definecolor{DarkOliveGreen}{rgb}{0.33,0.42,0.18}
\definecolor{DarkOrange1}{rgb}{1.00,0.50,0.00}
\definecolor{DarkOrange2}{rgb}{0.93,0.46,0.00}
\definecolor{DarkOrange3}{rgb}{0.80,0.40,0.00}
\definecolor{DarkOrange4}{rgb}{0.55,0.27,0.00}
\definecolor{DarkOrange}{rgb}{1.00,0.55,0.00}
\definecolor{DarkOrchid1}{rgb}{0.75,0.24,1.00}
\definecolor{DarkOrchid2}{rgb}{0.70,0.23,0.93}
\definecolor{DarkOrchid3}{rgb}{0.60,0.20,0.80}
\definecolor{DarkOrchid4}{rgb}{0.41,0.13,0.55}
\definecolor{DarkOrchid}{rgb}{0.60,0.20,0.80}
\definecolor{DarkRed}{rgb}{0.55,0.00,0.00}
\definecolor{DarkSalmon}{rgb}{0.91,0.59,0.48}
\definecolor{DarkSeaGreen1}{rgb}{0.76,1.00,0.76}
\definecolor{DarkSeaGreen2}{rgb}{0.71,0.93,0.71}
\definecolor{DarkSeaGreen3}{rgb}{0.61,0.80,0.61}
\definecolor{DarkSeaGreen4}{rgb}{0.41,0.55,0.41}
\definecolor{DarkSeaGreen}{rgb}{0.56,0.74,0.56}
\definecolor{DarkSlateBlue}{rgb}{0.28,0.24,0.55}
\definecolor{DarkSlateGray1}{rgb}{0.59,1.00,1.00}
\definecolor{DarkSlateGray2}{rgb}{0.55,0.93,0.93}
\definecolor{DarkSlateGray3}{rgb}{0.47,0.80,0.80}
\definecolor{DarkSlateGray4}{rgb}{0.32,0.55,0.55}
\definecolor{DarkSlateGray}{rgb}{0.18,0.31,0.31}
\definecolor{DarkSlateGrey}{rgb}{0.18,0.31,0.31}
\definecolor{DarkTurquoise}{rgb}{0.00,0.81,0.82}
\definecolor{DarkViolet}{rgb}{0.58,0.00,0.83}
\definecolor{DeepPink1}{rgb}{1.00,0.08,0.58}
\definecolor{DeepPink2}{rgb}{0.93,0.07,0.54}
\definecolor{DeepPink3}{rgb}{0.80,0.06,0.46}
\definecolor{DeepPink4}{rgb}{0.55,0.04,0.31}
\definecolor{DeepPink}{rgb}{1.00,0.08,0.58}
\definecolor{DeepSkyBlue1}{rgb}{0.00,0.75,1.00}
\definecolor{DeepSkyBlue2}{rgb}{0.00,0.70,0.93}
\definecolor{DeepSkyBlue3}{rgb}{0.00,0.60,0.80}
\definecolor{DeepSkyBlue4}{rgb}{0.00,0.41,0.55}
\definecolor{DeepSkyBlue}{rgb}{0.00,0.75,1.00}
\definecolor{DimGray}{rgb}{0.41,0.41,0.41}
\definecolor{DimGrey}{rgb}{0.41,0.41,0.41}
\definecolor{DodgerBlue1}{rgb}{0.12,0.56,1.00}
\definecolor{DodgerBlue2}{rgb}{0.11,0.53,0.93}
\definecolor{DodgerBlue3}{rgb}{0.09,0.45,0.80}
\definecolor{DodgerBlue4}{rgb}{0.06,0.31,0.55}
\definecolor{DodgerBlue}{rgb}{0.12,0.56,1.00}
\definecolor{FloralWhite}{rgb}{1.00,0.98,0.94}
\definecolor{ForestGreen}{rgb}{0.13,0.55,0.13}
\definecolor{GhostWhite}{rgb}{0.97,0.97,1.00}
\definecolor{GreenYellow}{rgb}{0.68,1.00,0.18}
\definecolor{HotPink1}{rgb}{1.00,0.43,0.71}
\definecolor{HotPink2}{rgb}{0.93,0.42,0.65}
\definecolor{HotPink3}{rgb}{0.80,0.38,0.56}
\definecolor{HotPink4}{rgb}{0.55,0.23,0.38}
\definecolor{HotPink}{rgb}{1.00,0.41,0.71}
\definecolor{IndianRed1}{rgb}{1.00,0.42,0.42}
\definecolor{IndianRed2}{rgb}{0.93,0.39,0.39}
\definecolor{IndianRed3}{rgb}{0.80,0.33,0.33}
\definecolor{IndianRed4}{rgb}{0.55,0.23,0.23}
\definecolor{IndianRed}{rgb}{0.80,0.36,0.36}
\definecolor{LavenderBlush1}{rgb}{1.00,0.94,0.96}
\definecolor{LavenderBlush2}{rgb}{0.93,0.88,0.90}
\definecolor{LavenderBlush3}{rgb}{0.80,0.76,0.77}
\definecolor{LavenderBlush4}{rgb}{0.55,0.51,0.53}
\definecolor{LavenderBlush}{rgb}{1.00,0.94,0.96}
\definecolor{LawnGreen}{rgb}{0.49,0.99,0.00}
\definecolor{LemonChiffon1}{rgb}{1.00,0.98,0.80}
\definecolor{LemonChiffon2}{rgb}{0.93,0.91,0.75}
\definecolor{LemonChiffon3}{rgb}{0.80,0.79,0.65}
\definecolor{LemonChiffon4}{rgb}{0.55,0.54,0.44}
\definecolor{LemonChiffon}{rgb}{1.00,0.98,0.80}
\definecolor{LightBlue1}{rgb}{0.75,0.94,1.00}
\definecolor{LightBlue2}{rgb}{0.70,0.87,0.93}
\definecolor{LightBlue3}{rgb}{0.60,0.75,0.80}
\definecolor{LightBlue4}{rgb}{0.41,0.51,0.55}
\definecolor{LightBlue}{rgb}{0.68,0.85,0.90}
\definecolor{LightCoral}{rgb}{0.94,0.50,0.50}
\definecolor{LightCyan1}{rgb}{0.88,1.00,1.00}
\definecolor{LightCyan2}{rgb}{0.82,0.93,0.93}
\definecolor{LightCyan3}{rgb}{0.71,0.80,0.80}
\definecolor{LightCyan4}{rgb}{0.48,0.55,0.55}
\definecolor{LightCyan}{rgb}{0.88,1.00,1.00}
\definecolor{LightGoldenrod1}{rgb}{1.00,0.93,0.55}
\definecolor{LightGoldenrod2}{rgb}{0.93,0.86,0.51}
\definecolor{LightGoldenrod3}{rgb}{0.80,0.75,0.44}
\definecolor{LightGoldenrod4}{rgb}{0.55,0.51,0.30}
\definecolor{LightGoldenrodYellow}{rgb}{0.98,0.98,0.82}
\definecolor{LightGoldenrod}{rgb}{0.93,0.87,0.51}
\definecolor{LightGray}{rgb}{0.83,0.83,0.83}
\definecolor{LightGreen}{rgb}{0.56,0.93,0.56}
\definecolor{LightGrey}{rgb}{0.83,0.83,0.83}
\definecolor{LightPink1}{rgb}{1.00,0.68,0.73}
\definecolor{LightPink2}{rgb}{0.93,0.64,0.68}
\definecolor{LightPink3}{rgb}{0.80,0.55,0.58}
\definecolor{LightPink4}{rgb}{0.55,0.37,0.40}
\definecolor{LightPink}{rgb}{1.00,0.71,0.76}
\definecolor{LightSalmon1}{rgb}{1.00,0.63,0.48}
\definecolor{LightSalmon2}{rgb}{0.93,0.58,0.45}
\definecolor{LightSalmon3}{rgb}{0.80,0.51,0.38}
\definecolor{LightSalmon4}{rgb}{0.55,0.34,0.26}
\definecolor{LightSalmon}{rgb}{1.00,0.63,0.48}
\definecolor{LightSeaGreen}{rgb}{0.13,0.70,0.67}
\definecolor{LightSkyBlue1}{rgb}{0.69,0.89,1.00}
\definecolor{LightSkyBlue2}{rgb}{0.64,0.83,0.93}
\definecolor{LightSkyBlue3}{rgb}{0.55,0.71,0.80}
\definecolor{LightSkyBlue4}{rgb}{0.38,0.48,0.55}
\definecolor{LightSkyBlue}{rgb}{0.53,0.81,0.98}
\definecolor{LightSlateBlue}{rgb}{0.52,0.44,1.00}
\definecolor{LightSlateGray}{rgb}{0.47,0.53,0.60}
\definecolor{LightSlateGrey}{rgb}{0.47,0.53,0.60}
\definecolor{LightSteelBlue1}{rgb}{0.79,0.88,1.00}
\definecolor{LightSteelBlue2}{rgb}{0.74,0.82,0.93}
\definecolor{LightSteelBlue3}{rgb}{0.64,0.71,0.80}
\definecolor{LightSteelBlue4}{rgb}{0.43,0.48,0.55}
\definecolor{LightSteelBlue}{rgb}{0.69,0.77,0.87}
\definecolor{LightYellow1}{rgb}{1.00,1.00,0.88}
\definecolor{LightYellow2}{rgb}{0.93,0.93,0.82}
\definecolor{LightYellow3}{rgb}{0.80,0.80,0.71}
\definecolor{LightYellow4}{rgb}{0.55,0.55,0.48}
\definecolor{LightYellow}{rgb}{1.00,1.00,0.88}
\definecolor{LimeGreen}{rgb}{0.20,0.80,0.20}
\definecolor{MediumAquamarine}{rgb}{0.40,0.80,0.67}
\definecolor{MediumBlue}{rgb}{0.00,0.00,0.80}
\definecolor{MediumOrchid1}{rgb}{0.88,0.40,1.00}
\definecolor{MediumOrchid2}{rgb}{0.82,0.37,0.93}
\definecolor{MediumOrchid3}{rgb}{0.71,0.32,0.80}
\definecolor{MediumOrchid4}{rgb}{0.48,0.22,0.55}
\definecolor{MediumOrchid}{rgb}{0.73,0.33,0.83}
\definecolor{MediumPurple1}{rgb}{0.67,0.51,1.00}
\definecolor{MediumPurple2}{rgb}{0.62,0.47,0.93}
\definecolor{MediumPurple3}{rgb}{0.54,0.41,0.80}
\definecolor{MediumPurple4}{rgb}{0.36,0.28,0.55}
\definecolor{MediumPurple}{rgb}{0.58,0.44,0.86}
\definecolor{MediumSeaGreen}{rgb}{0.24,0.70,0.44}
\definecolor{MediumSlateBlue}{rgb}{0.48,0.41,0.93}
\definecolor{MediumSpringGreen}{rgb}{0.00,0.98,0.60}
\definecolor{MediumTurquoise}{rgb}{0.28,0.82,0.80}
\definecolor{MediumVioletRed}{rgb}{0.78,0.08,0.52}
\definecolor{MidnightBlue}{rgb}{0.10,0.10,0.44}
\definecolor{MintCream}{rgb}{0.96,1.00,0.98}
\definecolor{MistyRose1}{rgb}{1.00,0.89,0.88}
\definecolor{MistyRose2}{rgb}{0.93,0.84,0.82}
\definecolor{MistyRose3}{rgb}{0.80,0.72,0.71}
\definecolor{MistyRose4}{rgb}{0.55,0.49,0.48}
\definecolor{MistyRose}{rgb}{1.00,0.89,0.88}
\definecolor{NavajoWhite1}{rgb}{1.00,0.87,0.68}
\definecolor{NavajoWhite2}{rgb}{0.93,0.81,0.63}
\definecolor{NavajoWhite3}{rgb}{0.80,0.70,0.55}
\definecolor{NavajoWhite4}{rgb}{0.55,0.47,0.37}
\definecolor{NavajoWhite}{rgb}{1.00,0.87,0.68}
\definecolor{NavyBlue}{rgb}{0.00,0.00,0.50}
\definecolor{OldLace}{rgb}{0.99,0.96,0.90}
\definecolor{OliveDrab1}{rgb}{0.75,1.00,0.24}
\definecolor{OliveDrab2}{rgb}{0.70,0.93,0.23}
\definecolor{OliveDrab3}{rgb}{0.60,0.80,0.20}
\definecolor{OliveDrab4}{rgb}{0.41,0.55,0.13}
\definecolor{OliveDrab}{rgb}{0.42,0.56,0.14}
\definecolor{OrangeRed1}{rgb}{1.00,0.27,0.00}
\definecolor{OrangeRed2}{rgb}{0.93,0.25,0.00}
\definecolor{OrangeRed3}{rgb}{0.80,0.22,0.00}
\definecolor{OrangeRed4}{rgb}{0.55,0.15,0.00}
\definecolor{OrangeRed}{rgb}{1.00,0.27,0.00}
\definecolor{PaleGoldenrod}{rgb}{0.93,0.91,0.67}
\definecolor{PaleGreen1}{rgb}{0.60,1.00,0.60}
\definecolor{PaleGreen2}{rgb}{0.56,0.93,0.56}
\definecolor{PaleGreen3}{rgb}{0.49,0.80,0.49}
\definecolor{PaleGreen4}{rgb}{0.33,0.55,0.33}
\definecolor{PaleGreen}{rgb}{0.60,0.98,0.60}
\definecolor{PaleTurquoise1}{rgb}{0.73,1.00,1.00}
\definecolor{PaleTurquoise2}{rgb}{0.68,0.93,0.93}
\definecolor{PaleTurquoise3}{rgb}{0.59,0.80,0.80}
\definecolor{PaleTurquoise4}{rgb}{0.40,0.55,0.55}
\definecolor{PaleTurquoise}{rgb}{0.69,0.93,0.93}
\definecolor{PaleVioletRed1}{rgb}{1.00,0.51,0.67}
\definecolor{PaleVioletRed2}{rgb}{0.93,0.47,0.62}
\definecolor{PaleVioletRed3}{rgb}{0.80,0.41,0.54}
\definecolor{PaleVioletRed4}{rgb}{0.55,0.28,0.36}
\definecolor{PaleVioletRed}{rgb}{0.86,0.44,0.58}
\definecolor{PapayaWhip}{rgb}{1.00,0.94,0.84}
\definecolor{PeachPuff1}{rgb}{1.00,0.85,0.73}
\definecolor{PeachPuff2}{rgb}{0.93,0.80,0.68}
\definecolor{PeachPuff3}{rgb}{0.80,0.69,0.58}
\definecolor{PeachPuff4}{rgb}{0.55,0.47,0.40}
\definecolor{PeachPuff}{rgb}{1.00,0.85,0.73}
\definecolor{PowderBlue}{rgb}{0.69,0.88,0.90}
\definecolor{RosyBrown1}{rgb}{1.00,0.76,0.76}
\definecolor{RosyBrown2}{rgb}{0.93,0.71,0.71}
\definecolor{RosyBrown3}{rgb}{0.80,0.61,0.61}
\definecolor{RosyBrown4}{rgb}{0.55,0.41,0.41}
\definecolor{RosyBrown}{rgb}{0.74,0.56,0.56}
\definecolor{RoyalBlue1}{rgb}{0.28,0.46,1.00}
\definecolor{RoyalBlue2}{rgb}{0.26,0.43,0.93}
\definecolor{RoyalBlue3}{rgb}{0.23,0.37,0.80}
\definecolor{RoyalBlue4}{rgb}{0.15,0.25,0.55}
\definecolor{RoyalBlue}{rgb}{0.25,0.41,0.88}
\definecolor{SaddleBrown}{rgb}{0.55,0.27,0.07}
\definecolor{SandyBrown}{rgb}{0.96,0.64,0.38}
\definecolor{SeaGreen1}{rgb}{0.33,1.00,0.62}
\definecolor{SeaGreen2}{rgb}{0.31,0.93,0.58}
\definecolor{SeaGreen3}{rgb}{0.26,0.80,0.50}
\definecolor{SeaGreen4}{rgb}{0.18,0.55,0.34}
\definecolor{SeaGreen}{rgb}{0.18,0.55,0.34}
\definecolor{SkyBlue1}{rgb}{0.53,0.81,1.00}
\definecolor{SkyBlue2}{rgb}{0.49,0.75,0.93}
\definecolor{SkyBlue3}{rgb}{0.42,0.65,0.80}
\definecolor{SkyBlue4}{rgb}{0.29,0.44,0.55}
\definecolor{SkyBlue}{rgb}{0.53,0.81,0.92}
\definecolor{SlateBlue1}{rgb}{0.51,0.44,1.00}
\definecolor{SlateBlue2}{rgb}{0.48,0.40,0.93}
\definecolor{SlateBlue3}{rgb}{0.41,0.35,0.80}
\definecolor{SlateBlue4}{rgb}{0.28,0.24,0.55}
\definecolor{SlateBlue}{rgb}{0.42,0.35,0.80}
\definecolor{SlateGray1}{rgb}{0.78,0.89,1.00}
\definecolor{SlateGray2}{rgb}{0.73,0.83,0.93}
\definecolor{SlateGray3}{rgb}{0.62,0.71,0.80}
\definecolor{SlateGray4}{rgb}{0.42,0.48,0.55}
\definecolor{SlateGray}{rgb}{0.44,0.50,0.56}
\definecolor{SlateGrey}{rgb}{0.44,0.50,0.56}
\definecolor{SpringGreen1}{rgb}{0.00,1.00,0.50}
\definecolor{SpringGreen2}{rgb}{0.00,0.93,0.46}
\definecolor{SpringGreen3}{rgb}{0.00,0.80,0.40}
\definecolor{SpringGreen4}{rgb}{0.00,0.55,0.27}
\definecolor{SpringGreen}{rgb}{0.00,1.00,0.50}
\definecolor{SteelBlue1}{rgb}{0.39,0.72,1.00}
\definecolor{SteelBlue2}{rgb}{0.36,0.67,0.93}
\definecolor{SteelBlue3}{rgb}{0.31,0.58,0.80}
\definecolor{SteelBlue4}{rgb}{0.21,0.39,0.55}
\definecolor{SteelBlue}{rgb}{0.27,0.51,0.71}
\definecolor{VioletRed1}{rgb}{1.00,0.24,0.59}
\definecolor{VioletRed2}{rgb}{0.93,0.23,0.55}
\definecolor{VioletRed3}{rgb}{0.80,0.20,0.47}
\definecolor{VioletRed4}{rgb}{0.55,0.13,0.32}
\definecolor{VioletRed}{rgb}{0.82,0.13,0.56}
\definecolor{WhiteSmoke}{rgb}{0.96,0.96,0.96}
\definecolor{YellowGreen}{rgb}{0.60,0.80,0.20}
\definecolor{aliceblue}{rgb}{0.94,0.97,1.00}
\definecolor{antiquewhite}{rgb}{0.98,0.92,0.84}
\definecolor{aquamarine1}{rgb}{0.50,1.00,0.83}
\definecolor{aquamarine2}{rgb}{0.46,0.93,0.78}
\definecolor{aquamarine3}{rgb}{0.40,0.80,0.67}
\definecolor{aquamarine4}{rgb}{0.27,0.55,0.45}
\definecolor{aquamarine}{rgb}{0.50,1.00,0.83}
\definecolor{azure1}{rgb}{0.94,1.00,1.00}
\definecolor{azure2}{rgb}{0.88,0.93,0.93}
\definecolor{azure3}{rgb}{0.76,0.80,0.80}
\definecolor{azure4}{rgb}{0.51,0.55,0.55}
\definecolor{azure}{rgb}{0.94,1.00,1.00}
\definecolor{beige}{rgb}{0.96,0.96,0.86}
\definecolor{bisque1}{rgb}{1.00,0.89,0.77}
\definecolor{bisque2}{rgb}{0.93,0.84,0.72}
\definecolor{bisque3}{rgb}{0.80,0.72,0.62}
\definecolor{bisque4}{rgb}{0.55,0.49,0.42}
\definecolor{bisque}{rgb}{1.00,0.89,0.77}
\definecolor{black}{rgb}{0.00,0.00,0.00}
\definecolor{blanchedalmond}{rgb}{1.00,0.92,0.80}
\definecolor{blue1}{rgb}{0.00,0.00,1.00}
\definecolor{blue2}{rgb}{0.00,0.00,0.93}
\definecolor{blue3}{rgb}{0.00,0.00,0.80}
\definecolor{blue4}{rgb}{0.00,0.00,0.55}
\definecolor{blueviolet}{rgb}{0.54,0.17,0.89}
\definecolor{blue}{rgb}{0.00,0.00,1.00}
\definecolor{brown1}{rgb}{1.00,0.25,0.25}
\definecolor{brown2}{rgb}{0.93,0.23,0.23}
\definecolor{brown3}{rgb}{0.80,0.20,0.20}
\definecolor{brown4}{rgb}{0.55,0.14,0.14}
\definecolor{brown}{rgb}{0.65,0.16,0.16}
\definecolor{burlywood1}{rgb}{1.00,0.83,0.61}
\definecolor{burlywood2}{rgb}{0.93,0.77,0.57}
\definecolor{burlywood3}{rgb}{0.80,0.67,0.49}
\definecolor{burlywood4}{rgb}{0.55,0.45,0.33}
\definecolor{burlywood}{rgb}{0.87,0.72,0.53}
\definecolor{cadetblue}{rgb}{0.37,0.62,0.63}
\definecolor{chartreuse1}{rgb}{0.50,1.00,0.00}
\definecolor{chartreuse2}{rgb}{0.46,0.93,0.00}
\definecolor{chartreuse3}{rgb}{0.40,0.80,0.00}
\definecolor{chartreuse4}{rgb}{0.27,0.55,0.00}
\definecolor{chartreuse}{rgb}{0.50,1.00,0.00}
\definecolor{chocolate1}{rgb}{1.00,0.50,0.14}
\definecolor{chocolate2}{rgb}{0.93,0.46,0.13}
\definecolor{chocolate3}{rgb}{0.80,0.40,0.11}
\definecolor{chocolate4}{rgb}{0.55,0.27,0.07}
\definecolor{chocolate}{rgb}{0.82,0.41,0.12}
\definecolor{coral1}{rgb}{1.00,0.45,0.34}
\definecolor{coral2}{rgb}{0.93,0.42,0.31}
\definecolor{coral3}{rgb}{0.80,0.36,0.27}
\definecolor{coral4}{rgb}{0.55,0.24,0.18}
\definecolor{coral}{rgb}{1.00,0.50,0.31}
\definecolor{cornflowerblue}{rgb}{0.39,0.58,0.93}
\definecolor{cornsilk1}{rgb}{1.00,0.97,0.86}
\definecolor{cornsilk2}{rgb}{0.93,0.91,0.80}
\definecolor{cornsilk3}{rgb}{0.80,0.78,0.69}
\definecolor{cornsilk4}{rgb}{0.55,0.53,0.47}
\definecolor{cornsilk}{rgb}{1.00,0.97,0.86}
\definecolor{cyan1}{rgb}{0.00,1.00,1.00}
\definecolor{cyan2}{rgb}{0.00,0.93,0.93}
\definecolor{cyan3}{rgb}{0.00,0.80,0.80}
\definecolor{cyan4}{rgb}{0.00,0.55,0.55}
\definecolor{cyan}{rgb}{0.00,1.00,1.00}
\definecolor{darkblue}{rgb}{0.00,0.00,0.55}
\definecolor{darkcyan}{rgb}{0.00,0.55,0.55}
\definecolor{darkgoldenrod}{rgb}{0.72,0.53,0.04}
\definecolor{darkgray}{rgb}{0.66,0.66,0.66}
\definecolor{darkgreen}{rgb}{0.00,0.39,0.00}
\definecolor{darkgrey}{rgb}{0.66,0.66,0.66}
\definecolor{darkkhaki}{rgb}{0.74,0.72,0.42}
\definecolor{darkmagenta}{rgb}{0.55,0.00,0.55}
\definecolor{darkolive}{rgb}{0.33,0.42,0.18}
\definecolor{darkorange}{rgb}{1.00,0.55,0.00}
\definecolor{darkorchid}{rgb}{0.60,0.20,0.80}
\definecolor{darkred}{rgb}{0.55,0.00,0.00}
\definecolor{darksalmon}{rgb}{0.91,0.59,0.48}
\definecolor{darksea}{rgb}{0.56,0.74,0.56}
\definecolor{darkslate}{rgb}{0.18,0.31,0.31}
\definecolor{darkslate}{rgb}{0.18,0.31,0.31}
\definecolor{darkslate}{rgb}{0.28,0.24,0.55}
\definecolor{darkturquoise}{rgb}{0.00,0.81,0.82}
\definecolor{darkviolet}{rgb}{0.58,0.00,0.83}
\definecolor{deeppink}{rgb}{1.00,0.08,0.58}
\definecolor{deepsky}{rgb}{0.00,0.75,1.00}
\definecolor{dimgray}{rgb}{0.41,0.41,0.41}
\definecolor{dimgrey}{rgb}{0.41,0.41,0.41}
\definecolor{dodgerblue}{rgb}{0.12,0.56,1.00}
\definecolor{firebrick1}{rgb}{1.00,0.19,0.19}
\definecolor{firebrick2}{rgb}{0.93,0.17,0.17}
\definecolor{firebrick3}{rgb}{0.80,0.15,0.15}
\definecolor{firebrick4}{rgb}{0.55,0.10,0.10}
\definecolor{firebrick}{rgb}{0.70,0.13,0.13}
\definecolor{floralwhite}{rgb}{1.00,0.98,0.94}
\definecolor{forestgreen}{rgb}{0.13,0.55,0.13}
\definecolor{gainsboro}{rgb}{0.86,0.86,0.86}
\definecolor{ghostwhite}{rgb}{0.97,0.97,1.00}
\definecolor{gold1}{rgb}{1.00,0.84,0.00}
\definecolor{gold2}{rgb}{0.93,0.79,0.00}
\definecolor{gold3}{rgb}{0.80,0.68,0.00}
\definecolor{gold4}{rgb}{0.55,0.46,0.00}
\definecolor{goldenrod1}{rgb}{1.00,0.76,0.15}
\definecolor{goldenrod2}{rgb}{0.93,0.71,0.13}
\definecolor{goldenrod3}{rgb}{0.80,0.61,0.11}
\definecolor{goldenrod4}{rgb}{0.55,0.41,0.08}
\definecolor{goldenrod}{rgb}{0.85,0.65,0.13}
\definecolor{gold}{rgb}{1.00,0.84,0.00}
\definecolor{gray0}{rgb}{0.00,0.00,0.00}
\definecolor{gray100}{rgb}{1.00,1.00,1.00}
\definecolor{gray10}{rgb}{0.10,0.10,0.10}
\definecolor{gray11}{rgb}{0.11,0.11,0.11}
\definecolor{gray12}{rgb}{0.12,0.12,0.12}
\definecolor{gray13}{rgb}{0.13,0.13,0.13}
\definecolor{gray14}{rgb}{0.14,0.14,0.14}
\definecolor{gray15}{rgb}{0.15,0.15,0.15}
\definecolor{gray16}{rgb}{0.16,0.16,0.16}
\definecolor{gray17}{rgb}{0.17,0.17,0.17}
\definecolor{gray18}{rgb}{0.18,0.18,0.18}
\definecolor{gray19}{rgb}{0.19,0.19,0.19}
\definecolor{gray1}{rgb}{0.01,0.01,0.01}
\definecolor{gray20}{rgb}{0.20,0.20,0.20}
\definecolor{gray21}{rgb}{0.21,0.21,0.21}
\definecolor{gray22}{rgb}{0.22,0.22,0.22}
\definecolor{gray23}{rgb}{0.23,0.23,0.23}
\definecolor{gray24}{rgb}{0.24,0.24,0.24}
\definecolor{gray25}{rgb}{0.25,0.25,0.25}
\definecolor{gray26}{rgb}{0.26,0.26,0.26}
\definecolor{gray27}{rgb}{0.27,0.27,0.27}
\definecolor{gray28}{rgb}{0.28,0.28,0.28}
\definecolor{gray29}{rgb}{0.29,0.29,0.29}
\definecolor{gray2}{rgb}{0.02,0.02,0.02}
\definecolor{gray30}{rgb}{0.30,0.30,0.30}
\definecolor{gray31}{rgb}{0.31,0.31,0.31}
\definecolor{gray32}{rgb}{0.32,0.32,0.32}
\definecolor{gray33}{rgb}{0.33,0.33,0.33}
\definecolor{gray34}{rgb}{0.34,0.34,0.34}
\definecolor{gray35}{rgb}{0.35,0.35,0.35}
\definecolor{gray36}{rgb}{0.36,0.36,0.36}
\definecolor{gray37}{rgb}{0.37,0.37,0.37}
\definecolor{gray38}{rgb}{0.38,0.38,0.38}
\definecolor{gray39}{rgb}{0.39,0.39,0.39}
\definecolor{gray3}{rgb}{0.03,0.03,0.03}
\definecolor{gray40}{rgb}{0.40,0.40,0.40}
\definecolor{gray41}{rgb}{0.41,0.41,0.41}
\definecolor{gray42}{rgb}{0.42,0.42,0.42}
\definecolor{gray43}{rgb}{0.43,0.43,0.43}
\definecolor{gray44}{rgb}{0.44,0.44,0.44}
\definecolor{gray45}{rgb}{0.45,0.45,0.45}
\definecolor{gray46}{rgb}{0.46,0.46,0.46}
\definecolor{gray47}{rgb}{0.47,0.47,0.47}
\definecolor{gray48}{rgb}{0.48,0.48,0.48}
\definecolor{gray49}{rgb}{0.49,0.49,0.49}
\definecolor{gray4}{rgb}{0.04,0.04,0.04}
\definecolor{gray50}{rgb}{0.50,0.50,0.50}
\definecolor{gray51}{rgb}{0.51,0.51,0.51}
\definecolor{gray52}{rgb}{0.52,0.52,0.52}
\definecolor{gray53}{rgb}{0.53,0.53,0.53}
\definecolor{gray54}{rgb}{0.54,0.54,0.54}
\definecolor{gray55}{rgb}{0.55,0.55,0.55}
\definecolor{gray56}{rgb}{0.56,0.56,0.56}
\definecolor{gray57}{rgb}{0.57,0.57,0.57}
\definecolor{gray58}{rgb}{0.58,0.58,0.58}
\definecolor{gray59}{rgb}{0.59,0.59,0.59}
\definecolor{gray5}{rgb}{0.05,0.05,0.05}
\definecolor{gray60}{rgb}{0.60,0.60,0.60}
\definecolor{gray61}{rgb}{0.61,0.61,0.61}
\definecolor{gray62}{rgb}{0.62,0.62,0.62}
\definecolor{gray63}{rgb}{0.63,0.63,0.63}
\definecolor{gray64}{rgb}{0.64,0.64,0.64}
\definecolor{gray65}{rgb}{0.65,0.65,0.65}
\definecolor{gray66}{rgb}{0.66,0.66,0.66}
\definecolor{gray67}{rgb}{0.67,0.67,0.67}
\definecolor{gray68}{rgb}{0.68,0.68,0.68}
\definecolor{gray69}{rgb}{0.69,0.69,0.69}
\definecolor{gray6}{rgb}{0.06,0.06,0.06}
\definecolor{gray70}{rgb}{0.70,0.70,0.70}
\definecolor{gray71}{rgb}{0.71,0.71,0.71}
\definecolor{gray72}{rgb}{0.72,0.72,0.72}
\definecolor{gray73}{rgb}{0.73,0.73,0.73}
\definecolor{gray74}{rgb}{0.74,0.74,0.74}
\definecolor{gray75}{rgb}{0.75,0.75,0.75}
\definecolor{gray76}{rgb}{0.76,0.76,0.76}
\definecolor{gray77}{rgb}{0.77,0.77,0.77}
\definecolor{gray78}{rgb}{0.78,0.78,0.78}
\definecolor{gray79}{rgb}{0.79,0.79,0.79}
\definecolor{gray7}{rgb}{0.07,0.07,0.07}
\definecolor{gray80}{rgb}{0.80,0.80,0.80}
\definecolor{gray81}{rgb}{0.81,0.81,0.81}
\definecolor{gray82}{rgb}{0.82,0.82,0.82}
\definecolor{gray83}{rgb}{0.83,0.83,0.83}
\definecolor{gray84}{rgb}{0.84,0.84,0.84}
\definecolor{gray85}{rgb}{0.85,0.85,0.85}
\definecolor{gray86}{rgb}{0.86,0.86,0.86}
\definecolor{gray87}{rgb}{0.87,0.87,0.87}
\definecolor{gray88}{rgb}{0.88,0.88,0.88}
\definecolor{gray89}{rgb}{0.89,0.89,0.89}
\definecolor{gray8}{rgb}{0.08,0.08,0.08}
\definecolor{gray90}{rgb}{0.90,0.90,0.90}
\definecolor{gray91}{rgb}{0.91,0.91,0.91}
\definecolor{gray92}{rgb}{0.92,0.92,0.92}
\definecolor{gray93}{rgb}{0.93,0.93,0.93}
\definecolor{gray94}{rgb}{0.94,0.94,0.94}
\definecolor{gray95}{rgb}{0.95,0.95,0.95}
\definecolor{gray96}{rgb}{0.96,0.96,0.96}
\definecolor{gray97}{rgb}{0.97,0.97,0.97}
\definecolor{gray98}{rgb}{0.98,0.98,0.98}
\definecolor{gray99}{rgb}{0.99,0.99,0.99}
\definecolor{gray9}{rgb}{0.09,0.09,0.09}
\definecolor{gray}{rgb}{0.75,0.75,0.75}
\definecolor{green1}{rgb}{0.00,1.00,0.00}
\definecolor{green2}{rgb}{0.00,0.93,0.00}
\definecolor{green3}{rgb}{0.00,0.80,0.00}
\definecolor{green4}{rgb}{0.00,0.55,0.00}
\definecolor{greenyellow}{rgb}{0.68,1.00,0.18}
\definecolor{green}{rgb}{0.00,1.00,0.00}
\definecolor{grey0}{rgb}{0.00,0.00,0.00}
\definecolor{grey100}{rgb}{1.00,1.00,1.00}
\definecolor{grey10}{rgb}{0.10,0.10,0.10}
\definecolor{grey11}{rgb}{0.11,0.11,0.11}
\definecolor{grey12}{rgb}{0.12,0.12,0.12}
\definecolor{grey13}{rgb}{0.13,0.13,0.13}
\definecolor{grey14}{rgb}{0.14,0.14,0.14}
\definecolor{grey15}{rgb}{0.15,0.15,0.15}
\definecolor{grey16}{rgb}{0.16,0.16,0.16}
\definecolor{grey17}{rgb}{0.17,0.17,0.17}
\definecolor{grey18}{rgb}{0.18,0.18,0.18}
\definecolor{grey19}{rgb}{0.19,0.19,0.19}
\definecolor{grey1}{rgb}{0.01,0.01,0.01}
\definecolor{grey20}{rgb}{0.20,0.20,0.20}
\definecolor{grey21}{rgb}{0.21,0.21,0.21}
\definecolor{grey22}{rgb}{0.22,0.22,0.22}
\definecolor{grey23}{rgb}{0.23,0.23,0.23}
\definecolor{grey24}{rgb}{0.24,0.24,0.24}
\definecolor{grey25}{rgb}{0.25,0.25,0.25}
\definecolor{grey26}{rgb}{0.26,0.26,0.26}
\definecolor{grey27}{rgb}{0.27,0.27,0.27}
\definecolor{grey28}{rgb}{0.28,0.28,0.28}
\definecolor{grey29}{rgb}{0.29,0.29,0.29}
\definecolor{grey2}{rgb}{0.02,0.02,0.02}
\definecolor{grey30}{rgb}{0.30,0.30,0.30}
\definecolor{grey31}{rgb}{0.31,0.31,0.31}
\definecolor{grey32}{rgb}{0.32,0.32,0.32}
\definecolor{grey33}{rgb}{0.33,0.33,0.33}
\definecolor{grey34}{rgb}{0.34,0.34,0.34}
\definecolor{grey35}{rgb}{0.35,0.35,0.35}
\definecolor{grey36}{rgb}{0.36,0.36,0.36}
\definecolor{grey37}{rgb}{0.37,0.37,0.37}
\definecolor{grey38}{rgb}{0.38,0.38,0.38}
\definecolor{grey39}{rgb}{0.39,0.39,0.39}
\definecolor{grey3}{rgb}{0.03,0.03,0.03}
\definecolor{grey40}{rgb}{0.40,0.40,0.40}
\definecolor{grey41}{rgb}{0.41,0.41,0.41}
\definecolor{grey42}{rgb}{0.42,0.42,0.42}
\definecolor{grey43}{rgb}{0.43,0.43,0.43}
\definecolor{grey44}{rgb}{0.44,0.44,0.44}
\definecolor{grey45}{rgb}{0.45,0.45,0.45}
\definecolor{grey46}{rgb}{0.46,0.46,0.46}
\definecolor{grey47}{rgb}{0.47,0.47,0.47}
\definecolor{grey48}{rgb}{0.48,0.48,0.48}
\definecolor{grey49}{rgb}{0.49,0.49,0.49}
\definecolor{grey4}{rgb}{0.04,0.04,0.04}
\definecolor{grey50}{rgb}{0.50,0.50,0.50}
\definecolor{grey51}{rgb}{0.51,0.51,0.51}
\definecolor{grey52}{rgb}{0.52,0.52,0.52}
\definecolor{grey53}{rgb}{0.53,0.53,0.53}
\definecolor{grey54}{rgb}{0.54,0.54,0.54}
\definecolor{grey55}{rgb}{0.55,0.55,0.55}
\definecolor{grey56}{rgb}{0.56,0.56,0.56}
\definecolor{grey57}{rgb}{0.57,0.57,0.57}
\definecolor{grey58}{rgb}{0.58,0.58,0.58}
\definecolor{grey59}{rgb}{0.59,0.59,0.59}
\definecolor{grey5}{rgb}{0.05,0.05,0.05}
\definecolor{grey60}{rgb}{0.60,0.60,0.60}
\definecolor{grey61}{rgb}{0.61,0.61,0.61}
\definecolor{grey62}{rgb}{0.62,0.62,0.62}
\definecolor{grey63}{rgb}{0.63,0.63,0.63}
\definecolor{grey64}{rgb}{0.64,0.64,0.64}
\definecolor{grey65}{rgb}{0.65,0.65,0.65}
\definecolor{grey66}{rgb}{0.66,0.66,0.66}
\definecolor{grey67}{rgb}{0.67,0.67,0.67}
\definecolor{grey68}{rgb}{0.68,0.68,0.68}
\definecolor{grey69}{rgb}{0.69,0.69,0.69}
\definecolor{grey6}{rgb}{0.06,0.06,0.06}
\definecolor{grey70}{rgb}{0.70,0.70,0.70}
\definecolor{grey71}{rgb}{0.71,0.71,0.71}
\definecolor{grey72}{rgb}{0.72,0.72,0.72}
\definecolor{grey73}{rgb}{0.73,0.73,0.73}
\definecolor{grey74}{rgb}{0.74,0.74,0.74}
\definecolor{grey75}{rgb}{0.75,0.75,0.75}
\definecolor{grey76}{rgb}{0.76,0.76,0.76}
\definecolor{grey77}{rgb}{0.77,0.77,0.77}
\definecolor{grey78}{rgb}{0.78,0.78,0.78}
\definecolor{grey79}{rgb}{0.79,0.79,0.79}
\definecolor{grey7}{rgb}{0.07,0.07,0.07}
\definecolor{grey80}{rgb}{0.80,0.80,0.80}
\definecolor{grey81}{rgb}{0.81,0.81,0.81}
\definecolor{grey82}{rgb}{0.82,0.82,0.82}
\definecolor{grey83}{rgb}{0.83,0.83,0.83}
\definecolor{grey84}{rgb}{0.84,0.84,0.84}
\definecolor{grey85}{rgb}{0.85,0.85,0.85}
\definecolor{grey86}{rgb}{0.86,0.86,0.86}
\definecolor{grey87}{rgb}{0.87,0.87,0.87}
\definecolor{grey88}{rgb}{0.88,0.88,0.88}
\definecolor{grey89}{rgb}{0.89,0.89,0.89}
\definecolor{grey8}{rgb}{0.08,0.08,0.08}
\definecolor{grey90}{rgb}{0.90,0.90,0.90}
\definecolor{grey91}{rgb}{0.91,0.91,0.91}
\definecolor{grey92}{rgb}{0.92,0.92,0.92}
\definecolor{grey93}{rgb}{0.93,0.93,0.93}
\definecolor{grey94}{rgb}{0.94,0.94,0.94}
\definecolor{grey95}{rgb}{0.95,0.95,0.95}
\definecolor{grey96}{rgb}{0.96,0.96,0.96}
\definecolor{grey97}{rgb}{0.97,0.97,0.97}
\definecolor{grey98}{rgb}{0.98,0.98,0.98}
\definecolor{grey99}{rgb}{0.99,0.99,0.99}
\definecolor{grey9}{rgb}{0.09,0.09,0.09}
\definecolor{grey}{rgb}{0.75,0.75,0.75}
\definecolor{honeydew1}{rgb}{0.94,1.00,0.94}
\definecolor{honeydew2}{rgb}{0.88,0.93,0.88}
\definecolor{honeydew3}{rgb}{0.76,0.80,0.76}
\definecolor{honeydew4}{rgb}{0.51,0.55,0.51}
\definecolor{honeydew}{rgb}{0.94,1.00,0.94}
\definecolor{hotpink}{rgb}{1.00,0.41,0.71}
\definecolor{indianred}{rgb}{0.80,0.36,0.36}
\definecolor{ivory1}{rgb}{1.00,1.00,0.94}
\definecolor{ivory2}{rgb}{0.93,0.93,0.88}
\definecolor{ivory3}{rgb}{0.80,0.80,0.76}
\definecolor{ivory4}{rgb}{0.55,0.55,0.51}
\definecolor{ivory}{rgb}{1.00,1.00,0.94}
\definecolor{khaki1}{rgb}{1.00,0.96,0.56}
\definecolor{khaki2}{rgb}{0.93,0.90,0.52}
\definecolor{khaki3}{rgb}{0.80,0.78,0.45}
\definecolor{khaki4}{rgb}{0.55,0.53,0.31}
\definecolor{khaki}{rgb}{0.94,0.90,0.55}
\definecolor{lavenderblush}{rgb}{1.00,0.94,0.96}
\definecolor{lavender}{rgb}{0.90,0.90,0.98}
\definecolor{lawngreen}{rgb}{0.49,0.99,0.00}
\definecolor{lemonchiffon}{rgb}{1.00,0.98,0.80}
\definecolor{lightblue}{rgb}{0.68,0.85,0.90}
\definecolor{lightcoral}{rgb}{0.94,0.50,0.50}
\definecolor{lightcyan}{rgb}{0.88,1.00,1.00}
\definecolor{lightgoldenrod}{rgb}{0.93,0.87,0.51}
\definecolor{lightgoldenrod}{rgb}{0.98,0.98,0.82}
\definecolor{lightgray}{rgb}{0.83,0.83,0.83}
\definecolor{lightgreen}{rgb}{0.56,0.93,0.56}
\definecolor{lightgrey}{rgb}{0.83,0.83,0.83}
\definecolor{lightpink}{rgb}{1.00,0.71,0.76}
\definecolor{lightsalmon}{rgb}{1.00,0.63,0.48}
\definecolor{lightsea}{rgb}{0.13,0.70,0.67}
\definecolor{lightsky}{rgb}{0.53,0.81,0.98}
\definecolor{lightslate}{rgb}{0.47,0.53,0.60}
\definecolor{lightslate}{rgb}{0.47,0.53,0.60}
\definecolor{lightslate}{rgb}{0.52,0.44,1.00}
\definecolor{lightsteel}{rgb}{0.69,0.77,0.87}
\definecolor{lightyellow}{rgb}{1.00,1.00,0.88}
\definecolor{limegreen}{rgb}{0.20,0.80,0.20}
\definecolor{linen}{rgb}{0.98,0.94,0.90}
\definecolor{magenta1}{rgb}{1.00,0.00,1.00}
\definecolor{magenta2}{rgb}{0.93,0.00,0.93}
\definecolor{magenta3}{rgb}{0.80,0.00,0.80}
\definecolor{magenta4}{rgb}{0.55,0.00,0.55}
\definecolor{magenta}{rgb}{1.00,0.00,1.00}
\definecolor{maroon1}{rgb}{1.00,0.20,0.70}
\definecolor{maroon2}{rgb}{0.93,0.19,0.65}
\definecolor{maroon3}{rgb}{0.80,0.16,0.56}
\definecolor{maroon4}{rgb}{0.55,0.11,0.38}
\definecolor{maroon}{rgb}{0.69,0.19,0.38}
\definecolor{mediumaquamarine}{rgb}{0.40,0.80,0.67}
\definecolor{mediumblue}{rgb}{0.00,0.00,0.80}
\definecolor{mediumorchid}{rgb}{0.73,0.33,0.83}
\definecolor{mediumpurple}{rgb}{0.58,0.44,0.86}
\definecolor{mediumsea}{rgb}{0.24,0.70,0.44}
\definecolor{mediumslate}{rgb}{0.48,0.41,0.93}
\definecolor{mediumspring}{rgb}{0.00,0.98,0.60}
\definecolor{mediumturquoise}{rgb}{0.28,0.82,0.80}
\definecolor{mediumviolet}{rgb}{0.78,0.08,0.52}
\definecolor{midnightblue}{rgb}{0.10,0.10,0.44}
\definecolor{mintcream}{rgb}{0.96,1.00,0.98}
\definecolor{mistyrose}{rgb}{1.00,0.89,0.88}
\definecolor{moccasin}{rgb}{1.00,0.89,0.71}
\definecolor{navajowhite}{rgb}{1.00,0.87,0.68}
\definecolor{navyblue}{rgb}{0.00,0.00,0.50}
\definecolor{navy}{rgb}{0.00,0.00,0.50}
\definecolor{oldlace}{rgb}{0.99,0.96,0.90}
\definecolor{olivedrab}{rgb}{0.42,0.56,0.14}
\definecolor{orange1}{rgb}{1.00,0.65,0.00}
\definecolor{orange2}{rgb}{0.93,0.60,0.00}
\definecolor{orange3}{rgb}{0.80,0.52,0.00}
\definecolor{orange4}{rgb}{0.55,0.35,0.00}
\definecolor{orangered}{rgb}{1.00,0.27,0.00}
\definecolor{orange}{rgb}{1.00,0.65,0.00}
\definecolor{orchid1}{rgb}{1.00,0.51,0.98}
\definecolor{orchid2}{rgb}{0.93,0.48,0.91}
\definecolor{orchid3}{rgb}{0.80,0.41,0.79}
\definecolor{orchid4}{rgb}{0.55,0.28,0.54}
\definecolor{orchid}{rgb}{0.85,0.44,0.84}
\definecolor{palegoldenrod}{rgb}{0.93,0.91,0.67}
\definecolor{palegreen}{rgb}{0.60,0.98,0.60}
\definecolor{paleturquoise}{rgb}{0.69,0.93,0.93}
\definecolor{paleviolet}{rgb}{0.86,0.44,0.58}
\definecolor{papayawhip}{rgb}{1.00,0.94,0.84}
\definecolor{peachpuff}{rgb}{1.00,0.85,0.73}
\definecolor{peru}{rgb}{0.80,0.52,0.25}
\definecolor{pink1}{rgb}{1.00,0.71,0.77}
\definecolor{pink2}{rgb}{0.93,0.66,0.72}
\definecolor{pink3}{rgb}{0.80,0.57,0.62}
\definecolor{pink4}{rgb}{0.55,0.39,0.42}
\definecolor{pink}{rgb}{1.00,0.75,0.80}
\definecolor{plum1}{rgb}{1.00,0.73,1.00}
\definecolor{plum2}{rgb}{0.93,0.68,0.93}
\definecolor{plum3}{rgb}{0.80,0.59,0.80}
\definecolor{plum4}{rgb}{0.55,0.40,0.55}
\definecolor{plum}{rgb}{0.87,0.63,0.87}
\definecolor{powderblue}{rgb}{0.69,0.88,0.90}
\definecolor{purple1}{rgb}{0.61,0.19,1.00}
\definecolor{purple2}{rgb}{0.57,0.17,0.93}
\definecolor{purple3}{rgb}{0.49,0.15,0.80}
\definecolor{purple4}{rgb}{0.33,0.10,0.55}
\definecolor{purple}{rgb}{0.63,0.13,0.94}
\definecolor{red1}{rgb}{1.00,0.00,0.00}
\definecolor{red2}{rgb}{0.93,0.00,0.00}
\definecolor{red3}{rgb}{0.80,0.00,0.00}
\definecolor{red4}{rgb}{0.55,0.00,0.00}
\definecolor{red}{rgb}{1.00,0.00,0.00}
\definecolor{rosybrown}{rgb}{0.74,0.56,0.56}
\definecolor{royalblue}{rgb}{0.25,0.41,0.88}
\definecolor{saddlebrown}{rgb}{0.55,0.27,0.07}
\definecolor{salmon1}{rgb}{1.00,0.55,0.41}
\definecolor{salmon2}{rgb}{0.93,0.51,0.38}
\definecolor{salmon3}{rgb}{0.80,0.44,0.33}
\definecolor{salmon4}{rgb}{0.55,0.30,0.22}
\definecolor{salmon}{rgb}{0.98,0.50,0.45}
\definecolor{sandybrown}{rgb}{0.96,0.64,0.38}
\definecolor{seagreen}{rgb}{0.18,0.55,0.34}
\definecolor{seashell1}{rgb}{1.00,0.96,0.93}
\definecolor{seashell2}{rgb}{0.93,0.90,0.87}
\definecolor{seashell3}{rgb}{0.80,0.77,0.75}
\definecolor{seashell4}{rgb}{0.55,0.53,0.51}
\definecolor{seashell}{rgb}{1.00,0.96,0.93}
\definecolor{sienna1}{rgb}{1.00,0.51,0.28}
\definecolor{sienna2}{rgb}{0.93,0.47,0.26}
\definecolor{sienna3}{rgb}{0.80,0.41,0.22}
\definecolor{sienna4}{rgb}{0.55,0.28,0.15}
\definecolor{sienna}{rgb}{0.63,0.32,0.18}
\definecolor{skyblue}{rgb}{0.53,0.81,0.92}
\definecolor{slateblue}{rgb}{0.42,0.35,0.80}
\definecolor{slategray}{rgb}{0.44,0.50,0.56}
\definecolor{slategrey}{rgb}{0.44,0.50,0.56}
\definecolor{snow1}{rgb}{1.00,0.98,0.98}
\definecolor{snow2}{rgb}{0.93,0.91,0.91}
\definecolor{snow3}{rgb}{0.80,0.79,0.79}
\definecolor{snow4}{rgb}{0.55,0.54,0.54}
\definecolor{snow}{rgb}{1.00,0.98,0.98}
\definecolor{springgreen}{rgb}{0.00,1.00,0.50}
\definecolor{steelblue}{rgb}{0.27,0.51,0.71}
\definecolor{tan1}{rgb}{1.00,0.65,0.31}
\definecolor{tan2}{rgb}{0.93,0.60,0.29}
\definecolor{tan3}{rgb}{0.80,0.52,0.25}
\definecolor{tan4}{rgb}{0.55,0.35,0.17}
\definecolor{tan}{rgb}{0.82,0.71,0.55}
\definecolor{thistle1}{rgb}{1.00,0.88,1.00}
\definecolor{thistle2}{rgb}{0.93,0.82,0.93}
\definecolor{thistle3}{rgb}{0.80,0.71,0.80}
\definecolor{thistle4}{rgb}{0.55,0.48,0.55}
\definecolor{thistle}{rgb}{0.85,0.75,0.85}
\definecolor{tomato1}{rgb}{1.00,0.39,0.28}
\definecolor{tomato2}{rgb}{0.93,0.36,0.26}
\definecolor{tomato3}{rgb}{0.80,0.31,0.22}
\definecolor{tomato4}{rgb}{0.55,0.21,0.15}
\definecolor{tomato}{rgb}{1.00,0.39,0.28}
\definecolor{turquoise1}{rgb}{0.00,0.96,1.00}
\definecolor{turquoise2}{rgb}{0.00,0.90,0.93}
\definecolor{turquoise3}{rgb}{0.00,0.77,0.80}
\definecolor{turquoise4}{rgb}{0.00,0.53,0.55}
\definecolor{turquoise}{rgb}{0.25,0.88,0.82}
\definecolor{violetred}{rgb}{0.82,0.13,0.56}
\definecolor{violet}{rgb}{0.93,0.51,0.93}
\definecolor{wheat1}{rgb}{1.00,0.91,0.73}
\definecolor{wheat2}{rgb}{0.93,0.85,0.68}
\definecolor{wheat3}{rgb}{0.80,0.73,0.59}
\definecolor{wheat4}{rgb}{0.55,0.49,0.40}
\definecolor{wheat}{rgb}{0.96,0.87,0.70}
\definecolor{whitesmoke}{rgb}{0.96,0.96,0.96}
\definecolor{white}{rgb}{1.00,1.00,1.00}
\definecolor{yellow1}{rgb}{1.00,1.00,0.00}
\definecolor{yellow2}{rgb}{0.93,0.93,0.00}
\definecolor{yellow3}{rgb}{0.80,0.80,0.00}
\definecolor{yellow4}{rgb}{0.55,0.55,0.00}
\definecolor{yellowgreen}{rgb}{0.60,0.80,0.20}
\definecolor{yellow}{rgb}{1.00,1.00,0.00}

\title[Star formation in red mergers]
    {{\color{black}Recent star formation in local, morphologically disturbed spheroidal galaxies on the optical red sequence}}

\author[Sugata Kaviraj]
{Sugata Kaviraj\thanks{E-mail: s.kaviraj@imperial.ac.uk}$^{1,2,3}$\\
$^1$Blackett Laboratory, Imperial College London, London SW7 2AZ, UK\\
$^2$Mullard Space Science Laboratory, Holmbury St. Mary, Dorking,
Surrey RH5 6NT UK\\
$^3$Department of Physics, University of Oxford, Keble Road,
Oxford OX1 3RH, UK}

\begin{document}

\date{\today}

\pagerange{\pageref{firstpage}--\pageref{lastpage}} \pubyear{2009}

\maketitle

\label{firstpage}


\begin{abstract}
{\color{black}We combine GALEX (ultra-violet; UV) and SDSS
(optical) photometry to study the recent star formation histories
of $\sim100$ field galaxies on the optical red sequence, a large
fraction of which exhibit widespread signs of disturbed
morphologies in deep optical imaging that are consistent with
recent merging events. More than 70\% of bulge-dominated galaxies
in this sample show tidal features at a surface brightness limit
of $\mu\sim28$ mag arcsec$^{-2}$. We find that, while they inhabit
the \emph{optical} red sequence, they show a wide spread in their
UV colours ($\sim$ 4 mags), akin to what has been discovered
recently in the general early-type population. {\color{black}A
strong correlation is found between UV colour and the strength of
the tidal distortions, such that the bluest galaxies are more
distorted.} This strongly suggests that the blue UV colours seen
in many nearby early-types are driven by (low-level)
merger-induced star formation within the last 3 Gyrs, contributing
less than 10\% of the stellar mass. If the \emph{ongoing} mergers
in this sample, which have a median mass ratio of 1:4, are
representative of the nearby red merger population, then less than
$\sim25$\% of the new stellar mass in the remnants is typically
added through merger-induced star formation. While the dust
extinction in the inter-stellar medium (ISM) in these galaxies is
small ($E^{ISM}_{B-V}<0.1$), the local dust content of the
star-forming regions is, on average, a factor of $\sim3$ higher.
Finally, we use our theoretical machinery to provide a recipe for
calculating the age of the most recent star formation event
($t_2$) in nearby ($z\lesssim0.1$) red early-type galaxies: $\log
t_2 {\color{black}\hspace{0.05in}\textnormal{[Gyrs]}} \sim
0.6^{\pm0.03}\times[(NUV-u)-(g-z)-1.73^{\pm0.03}]$, where
$NUV,u,g$ and $z$ are the observed photometric magnitudes of the
galaxies in the GALEX/SDSS filtersets.}
\end{abstract}




\section{Introduction}
The formation histories of luminous, spheroidal galaxies have
posed a serious challenge to models of galaxy formation for the
last few decades. The classical `monolithic collapse' hypothesis,
that followed the model of \cite{ELS62} for the formation of the
Galaxy \cite[e.g.][]{Larson74,Chiosi2002}, postulated that stellar
populations in early-type galaxies form in short, highly efficient
starbursts at high redshift ($z\gg1$) and evolve purely passively
thereafter. Indeed, the \emph{optical} properties of the
early-type population and their strict obedience to simple scaling
relations are remarkably consistent with such a simple formation
scenario. The small scatter in the early-type `Fundamental Plane'
\citep[e.g.][]{Jorg1996,Saglia1997} and its apparent lack of
evolution with look-back time
\citep[e.g.][]{Forbes1998,Peebles2002,Franx1993,Franx1995,VD1996},
the homogeneity and lack of redshift evolution in their optical
colours
\citep[e.g.][]{BLE92,Bender1997,Ellis97,Stanford98,Gladders98,VD2000}
and evidence for short ($<1$ Gyr) star formation timescales in
these systems, deduced from the over-abundance of $\alpha$
elements \citep[e.g.][]{Thomas1999}, all suggest that the bulk of
the stellar population in early-type galaxies does indeed form at
high redshift ($z>2$).

Nevertheless, consensus, in recent years, has moved away from
monolithic collapse towards a more gradual assembly of these
systems in the framework of the standard $\Lambda$CDM galaxy
formation paradigm. Following the seminal work of
\citet{Toomre_mergers}, numerical simulations have consistently
demonstrated that galaxy collisions end in rapid merging - a
fundamental feature of the standard model - and typically produce
spheroidal remnants \citep[e.g.][]{Barnes1992a} in cases where the
mass ratios of the progenitors are large ($\geq1:3$).
{\color{black}Semi-analytical models of galaxy formation, that
incorporate these results, have had reasonable success in
reproducing the morphological mix of the Universe and the
(photometric) properties of the early-type population
\citep[e.g.][]{Cole2000,Hatton2003,Kaviraj2005a,deLucia2006}}. The
predicted star formation histories (SFHs) of early-type galaxies
(at least in clusters) in the semi-analytical framework have been
shown to be \emph{quasi-monolithic}, resulting in good agreement
with the optical colours and their evolution with redshift
\citep[][]{Kaviraj2005a}. As a result, the observed optical
properties of the early-type population and their consistency with
the expectations of monolithic collapse are \emph{not} proof of a
uniquely monolithic origin.

A drawback of optical data is its relative insensitivity to small
amounts of {\color{black}recent star formation (RSF)}. The optical
spectrum remains largely unaffected by the minority of stellar
mass that is expected to form in these systems at low and
intermediate redshifts ($z<1$), which makes it difficult to
accurately measure early-type star formation histories (SFHs) over
the last 8 billion years.  However, while it does not impact the
optical spectrum (within typical observational and theoretical
uncertainties), a small mass fraction (a few percent) of young (a
few tens of Myrs old) stars strongly affects the rest-frame UV
spectrum shortward of 3000\AA. Furthermore, the UV remains largely
unaffected by the age-metallicity degeneracy \citep{Worthey1994}
that plagues optical analyses (Kaviraj et al. 2007a), making it an
ideal photometric indicator of RSF. The advent of the GALEX space
telescope (Martin et al. 2005) has provided us with an
unprecedented opportunity to harness the sensitivity of the UV to
young stars to \emph{quantify} the presence of RSF in local
early-type galaxies. Following the early work of \citet{Yi2005},
Kaviraj et al. (2007b; K07 hereafter), Schawinski et al. (2007a)
and Kaviraj et al. (2008) have carefully studied the UV properties
of the early-type population at late epochs (see
\citet{Kaviraj_etg_review} for a review), to convert the UV flux
observed in these systems into a measurement of the RSF, taking
into account potential contributions to the UV spectrum from old,
evolved stellar stages such as horizontal branch (HB) stars.
{\color{black}It is worth noting that models where early-types
host only old stars with realistic metallicity distributions that
satisfy both the mean metallicity and alpha-enhancements of nearby
early-types clearly require RSF (see Section 6.2 in K07)}. These
results indicate that early-types of all luminosities form stars
over the lifetime of the Universe, with massive systems
accumulating 10-15\% of their stars since $z\sim1$, plausibly
through minor mergers with small, gas-rich companions
\citep{Kaviraj2009}.

{\color{black}While such recent efforts have made a compelling
case for RSF-driven UV excess in the early-type population, we
should note some caveats to this analysis. The contamination of
the UV spectrum by old HB stars is a possibility at low redshift,
especially in giant elliptical galaxies \citep[see
e.g.][]{Yi97,Yi99}. However, studies of the rest-frame UV at high
redshift ($z>0.5$), where the HB is absent, reveal a similarly
large scatter in the rest-frame UV colour-magnitude relation (CMR)
to what is observed in the nearby Universe, strongly suggesting
that the principal drivers of the UV excess are not HB stars
(Kaviraj et al. 2008). It is conceivable that there is some
contribution to the UV flux from alternative hot stellar
populations such as blue stragglers - stars found above the
turn-off in Hertzprung-Russell diagrams \citep{Wheeler1979} -
typically within old globular clusters (GCs). Since they are
likely to form through stellar interactions \citep{Mapelli2006},
these objects tend to inhabit the cores of GCs, where the stellar
density is high. However, GCs (to which their constituent blue
stragglers contribute a vanishing fraction of stellar mass) host a
small fraction of the total stellar mass in the galaxy, making it
unlikely that the blue straggler population are responsible for
the UV flux in elliptical galaxies \citep{Mochkovitch1986}.

We note that such alterative contributions to the UV flux in
early-type galaxies have not been fully explored in the context of
contemporary rest-frame UV data from GALEX at low redshift and
deep optical surveys at high redshift. While RSF provides the most
attractive solution, both in terms of the observed UV spectra of
individual galaxies and in the context of the currently accepted
galaxy formation paradigm, the (presumably minor) role of these
alternate sources of UV flux remains an open question. A
potentially robust discriminant between RSF and old stars is the
\emph{spatial distribution} of UV flux in the galaxy. If it is
dominated by old stars (e.g. HB stars), the UV light is expected
to be a smooth background, perhaps with a colour gradient (due to
metallicity variations across the galaxy), but without large
pixel-to-pixel variations or pronounced fine structure. A firm
understanding of the origin of the UV flux would greatly benefit
from high-resolution imaging of nearby early-types, e.g. using
forthcoming instruments such as the HST-WFC3 camera.}

{\color{black}In this work, we study the GALEX (UV) and SDSS
(optical) properties of early-type galaxies on the optical red
sequence, originally studied by \citet[][vD05 hereafter]{VD2005}.}
The novelty of the vD05 study was the unprecedented depth of the
optical imaging employed to study the red galaxy population.
{\color{black}Using $\sim$27,000 second exposures on 4m class
telescopes (equivalent to $\sim$20 hrs on a 2.5m class telescope
like the one used for the SDSS) vD05 found widespread signatures
of recent merging in a large fraction of their early-type sample -
more than 70\% of bulge-dominated (E/S0) galaxies on the optical
red sequence exhibit distorted morphologies, with predominantly
red tidal features that extend to spatial scales $>50$ kpc.} The
deep images allow a \emph{quantitative} measure of the perturbed
morphologies of nearby early-type systems, which is not possible
with the substantially shallower imaging offered by other surveys
on 1-2.5m telescopes. As an example, the effective exposure time
of the SDSS - on which most of the recent early-type efforts are
based - is only $\sim$51s in each filter (York et al. 2000).

GALEX UV imaging of the vD05 sample provides a unique opportunity
to study the star formation history (SFH) of systems that are, to
high confidence, recent merger remnants. We first establish that
the vD05 sample is representative of the general luminous red
galaxy population at similar redshifts. We then combine the
GALEX/SDSS photometry of the vD05 sample to quantify the RSF in
these systems and explore the correlations between $(NUV-r)$, the
derived SFHs and the strength of the tidal distortions.
{\color{black}Finally, we use our theoretical machinery to derive
a simple recipe to translate the observed NUV and optical colours
of local early-type galaxies into an estimate of the age of the
last star formation episode in these galaxies.}


{\color{black}\section{The galaxy sample}} The original red galaxy
sample of vD05 was selected from two deep extragalactic surveys:
the Multi-wavelength Survey by Yale-Chile {\color{black}(MUSYC;
Gawiser et al. 2006)} and the NOAO Deep Wide-Field Survey
\citep[NDWFS;][]{Jannuzi1999}. The combined dataset covers 10.5
deg$^2$ of sky, reaching 1$\sigma$ surface brightness levels of
$\mu\sim29$ mag arcsec$^{-2}$. We refer readers to Section 3 of
vD05 for a comprehensive description of the dataset.

For the purposes of this study the vD05 sample was cross-matched
with both the SDSS DR4 and GALEX GR2.\footnote{{\color{black}The
positional matching with GALEX was performed within a 4" radius -
the fiducial GALEX PSF is 6"}}. Out of the original 126 galaxies
in the vD05 sample, the cross-matching process yields 101 galaxies
with both GALEX and SDSS photometry. The galaxies in this sample
have a mean redshift of $z\sim0.11$. The early-type galaxies, that
form the focus of this study, typically have super-L* luminosities
($M_r^* \sim -21.15$ in the nearby Universe; see Bernardi et al.
2003) and their masses range between $10^{10.4}$ and $10^{11.6}$
M$_{\odot}$.

{\color{black}Finally, we note that vD05 provides a quantitative
measure of the tidal distortions in each object by comparing
galaxy images to elliptical galaxy models, generated using the
\emph{ellipse} task in IRAF (see Section 5.3 in vD05 for a
description of the procedure). This tidal parameter (denoted by
$T$ in this paper) measures the median absolute deviation of the
(fractional) residuals from the model fit. Comparison of the value
of $T$ to visual inspection of the galaxy images indicates that,
$T\leq0.08$ in galaxies where the distortions are not obvious to
the eye, $0.13\leq T\leq0.19$ for weakly disturbed systems and
$T\geq0.19$ for systems that show strong tidal distortion. The
maximum value of the tidal parameter in this sample is $\sim0.47$.
In the following sections, we use this tidal parameter to
represent the extent of the morphological disturbance in each
galaxy.}


\vspace{0.3in} {\color{black}\section{Emission line properties,
spatial distributions and UV-optical colours}}

{\color{black}\subsection{Emission line properties: AGN and
star-formation activity}} The SDSS spectra of individual galaxies
were used to determine the presence of Type II AGN from optical
emission line ratios, using a `BPT'-type (e.g. Baldwin et al.
1981) analysis.

\begin{figure}
\begin{center}
\includegraphics[width=3.5in]{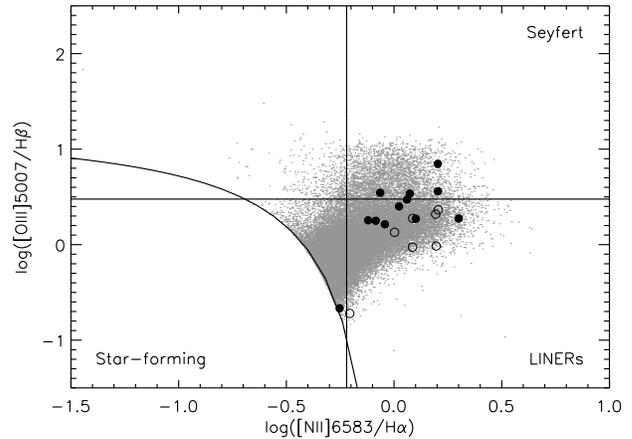}
\caption{{\color{black}A `BPT'-type plot for the vD05 red galaxy
sample, which indicates that only two galaxies lie near the star
forming sequence and that the bulk of the emission-line activity
is LINER-like.} The grey dots indicate the AGN population in the
low-redshift Universe, studied by Kauffmann et al. (2003). The
large circles denote galaxies in the vD05 sample. Galaxies are
shown using filled circles if all four lines have $S/N>3$.
Galaxies where only the [NII] and H$\alpha$ are detected are shown
using open circles.} \label{fig:bpt}
\end{center}
\end{figure}

In Figure \ref{fig:bpt} we present the BPT plot for the vD05
sample. Grey dots indicate the AGN population in the low-redshift
Universe, extracted by Kauffmann et al. (2003). Large circles
denote galaxies in the vD05 sample. Galaxies are shown using
filled circles if all four emission lines have $S/N>3$. Galaxies
where only [NII] and H$\alpha$ are detected are shown using open
circles. Note that while the detection of [NII] and H$\alpha$
allows us to classify these galaxies as AGN, the \emph{type} of
the AGN remains unclear, since the position on the y-axis depends
on the strength of [OIII] and H$\beta$ which are not robustly
detected.

{\color{black}It is remarkable that only two of the galaxies lie
close to the `star-forming' sequence of Kauffmann et al (which is
demarcated by the solid curve in the BPT plot)}. This is strong
evidence that the red mergers in vD05 are indeed mostly `dry' and
are not accompanied by significant ongoing star formation
\citep[see also][]{Whitaker2008}. The AGN fraction in this sample
is $\sim11$\%, consistent within errors ($\sim\pm3$\%) with the
fraction  found in the general early-type population (15\%) drawn
from the entire SDSS at low redshift
\citep[e.g.][]{Schawinski2007b}. It is also worth noting that most
of the AGN in this sample are classified as LINERs. A significant
fraction of these systems could, in fact, be systems with
declining star formation, where the `LINER-like' emission is
driven by post-AGB stars and white dwarfs
\citep[see][]{Stasinska2008}. While further exploration of this
issue is beyond the scope of this paper, we do not find compelling
evidence of strong AGN activity in the vD05 sample.


\vspace{0.3in}
\subsection{Spatial distributions and UV-optical colours} 
\begin{figure}
$\begin{array}{c}
\includegraphics[width=3.5in]{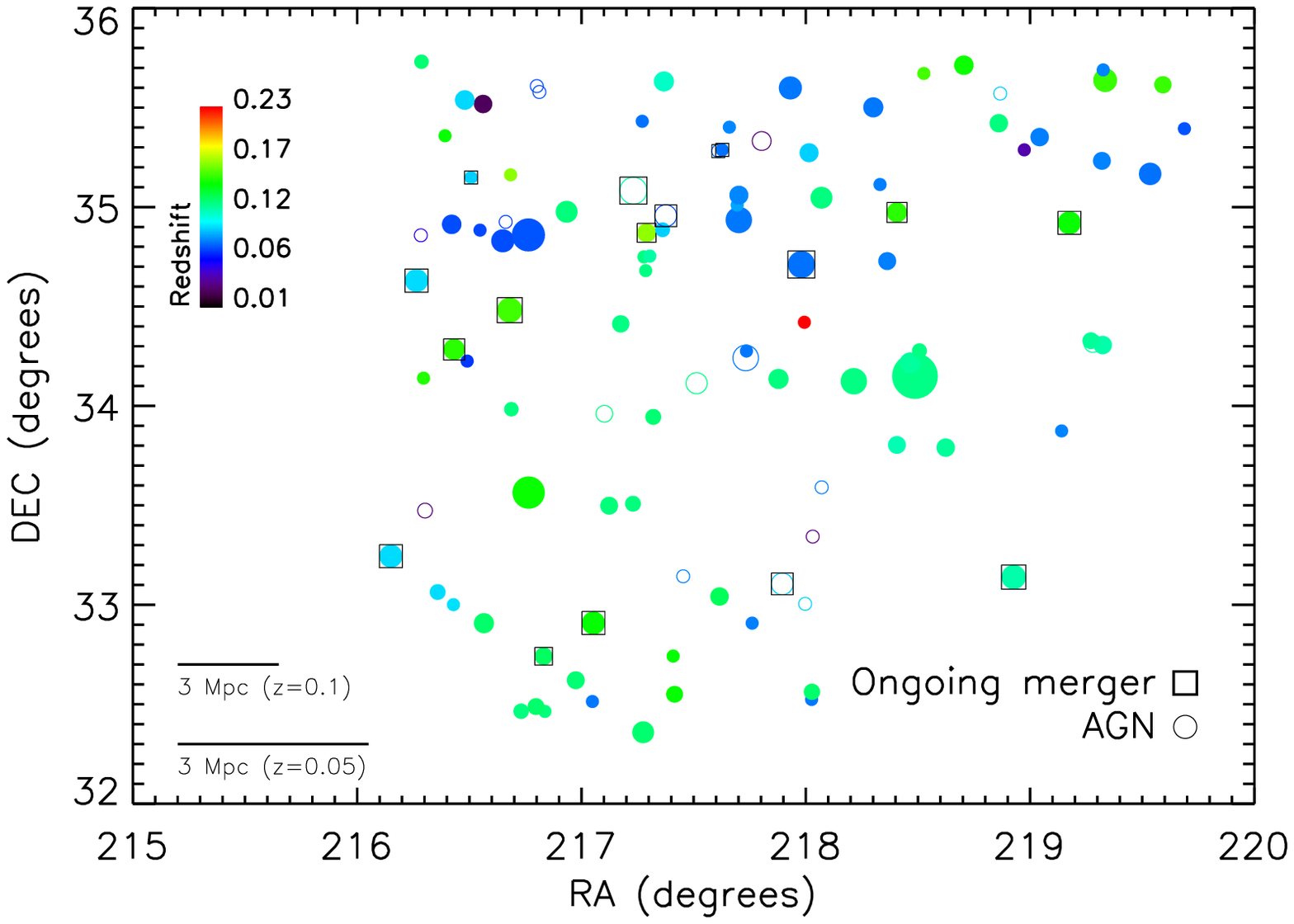}\\
\includegraphics[width=3.5in]{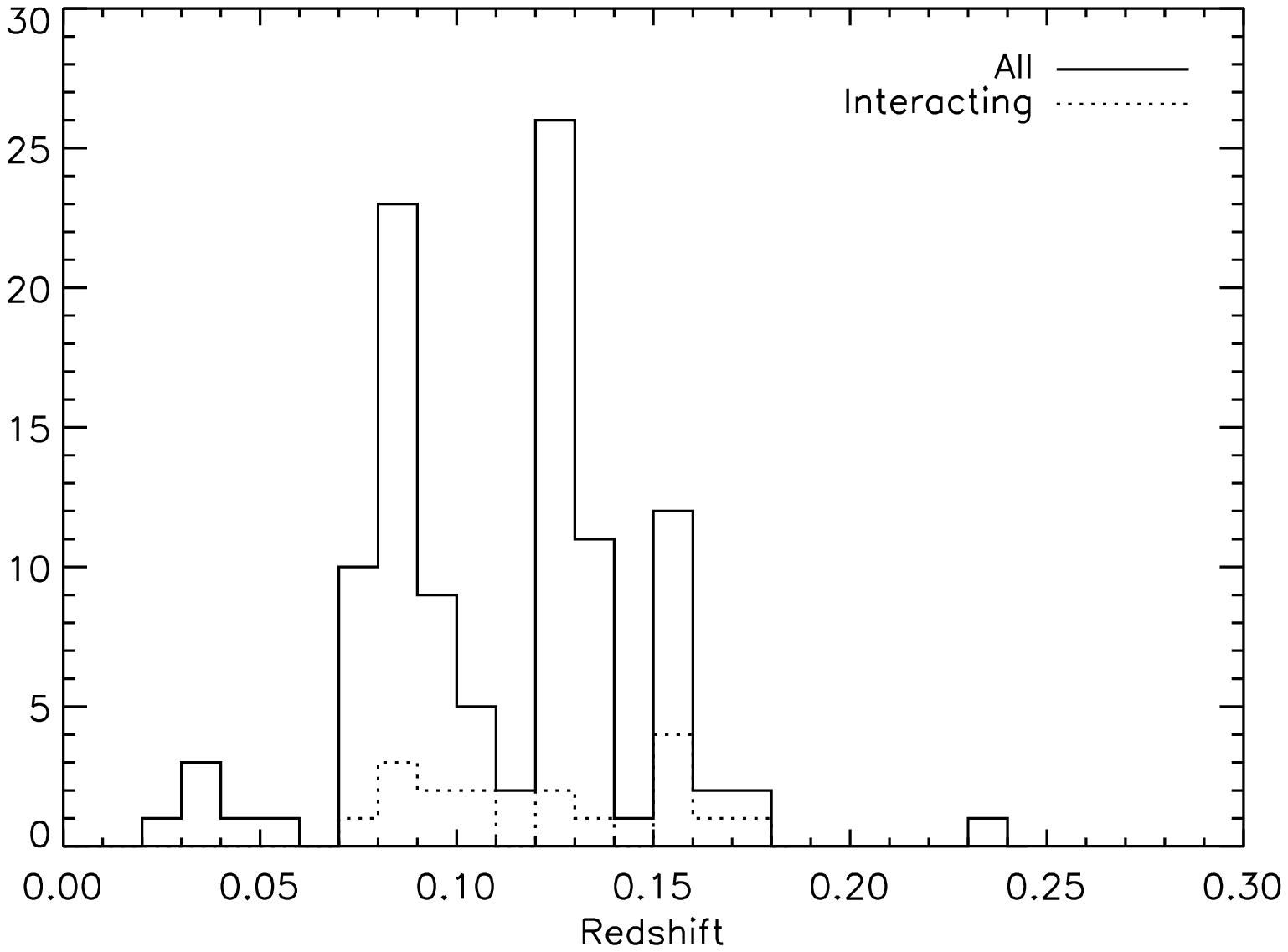}
\end{array}$
\caption{TOP: Spatial coordinates (RA, DEC and redshift) of the
galaxy sample studied here. BOTTOM: Comparison of the redshift
distribution of the general galaxy sample to that of only the
interacting population. {\color{black}Our results indicate that
the vD05 galaxies are generally in the `field'.} Symbol sizes are
proportional to the tidality parameter described in Section 2.
Galaxies with active AGN are shown using open symbols and objects
that are currently interacting are indicated using boxes.
{\color{black}Open boxes indicate ongoing mergers with AGN
activity.} We also indicate the angular sizes, at $z=0.1$ and
$z=0.05$, of a 3 Mpc structure (similar to the extent of the Coma
cluster). The redshift binning in this plot is $\delta z\sim0.01$,
which translates to a distance of $\sim$50 Mpc at $z=0.1$ in the
standard cosmology.} \label{fig:spatial_coordinates}
\end{figure}

We begin
by studying some basic properties of the vD05 objects. Figure
\ref{fig:spatial_coordinates} indicates the spatial coordinates
(RA, DEC and redshift) of the galaxy sample (top) and its redshift
distribution (bottom). Symbol sizes are proportional to the tidal
parameter described in Section 2. Galaxies that contain active AGN
are shown using open symbols and objects that are currently
interacting are indicated using boxes. The redshifts of the
galaxies in the top panel of this figure are shown colour-coded.
We also indicate the angular sizes, at $z=0.1$ and $z=0.05$, of a
3 Mpc structure (similar to the extent of the Coma cluster).

Given the high incidence of tidal features in the vD05 early-type
sample, we first check the spatial distribution of the objects to
look for particular overdensities that might lead to the observed
high frequency of interactions in the sample. The vD05 study did
not consider the spatial distribution of the galaxy sample due to
a lack of redshifts. They did estimate, however, that the median
redshift of the galaxies was $z\sim0.1$, based on the optical
($B-R$) colour of an $L^*$ elliptical galaxy. Based on the
spectroscopic redshifts from the SDSS (bottom panel of Figure
\ref{fig:spatial_coordinates}) we find that the mean redshift of
this sample to be $\sim0.11$, in good agreement with the value
estimated by vD05.

We find that the sample occupies three main redshift peaks. Note,
however, that the redshift binning used in Figure
\ref{fig:spatial_coordinates} is $\delta z=0.01$ - which
translates to a distance of $\sim$50 Mpc at $z=0.1$ in the
standard cosmology - and that the peaks have widths of at least
$\delta z=0.02$. Given the high accuracy of the SDSS redshifts
($\sim10^{-4}$), the redshift distribution does not indicate that
the galaxies are part of structures as compact as cluster cores
(which are sub-Mpc structures). Furthermore, the transverse
distances (in RA and DEC) are also quite large compared to, for
example, the full extent of the Coma cluster (3 Mpc). Given that
the central regions of nearby clusters are dominated by red
galaxies \citep{Dressler80}, we do not find compelling evidence
for the vD05 sample to be drawn out of a small number of overdense
regions. The galaxies therefore appear to be, on average, in the
`field'.

\begin{figure}
$\begin{array}{c}
\includegraphics[width=3.5in]{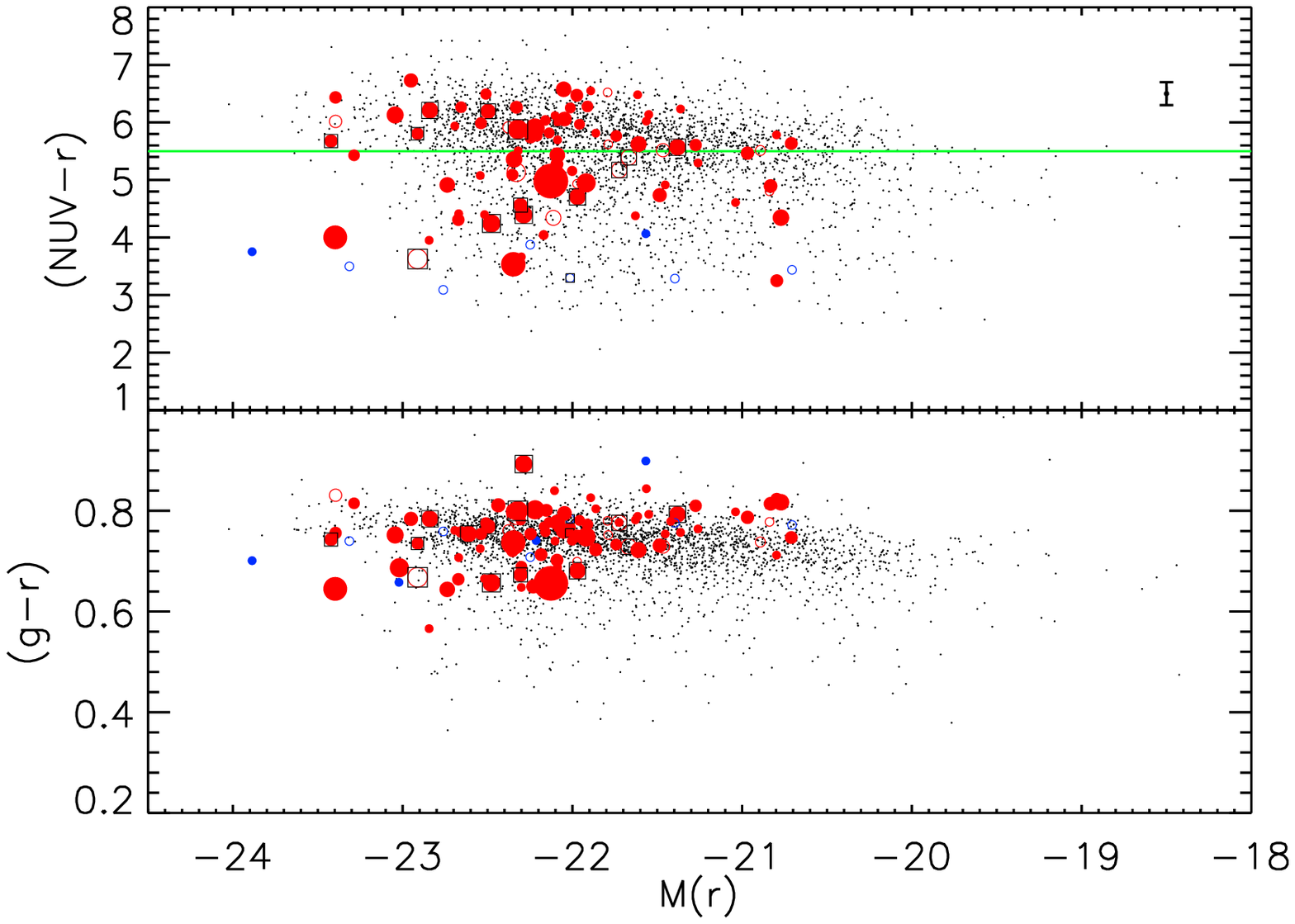}\\
\includegraphics[width=3.5in]{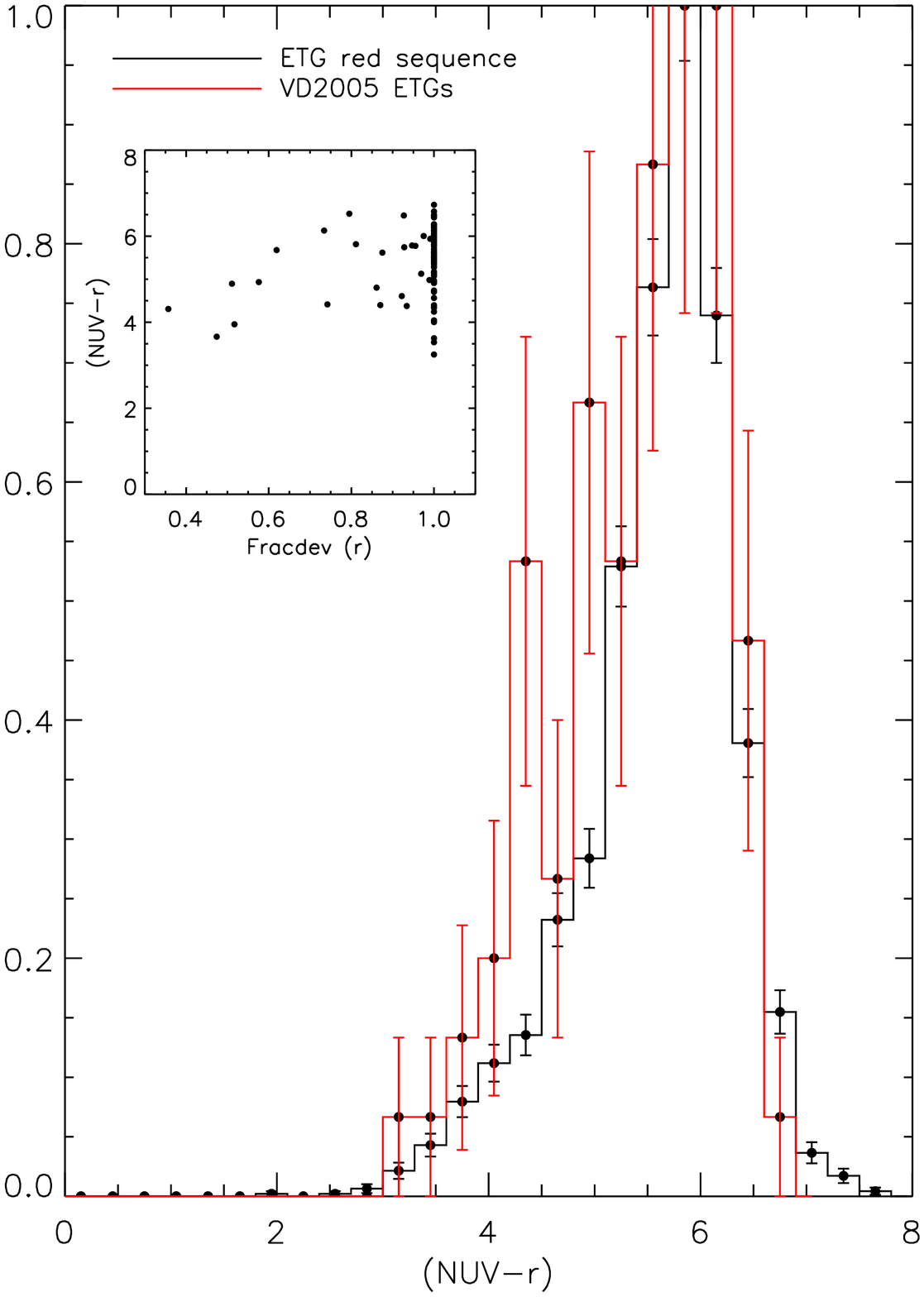}
\end{array}$
\caption{TOP: The optical and UV colour-magnitude relations
(CMRs).{\color{black} While selected to be optically red, the vD05
sample has a large range of UV colours}. Red symbols indicate
early-type galaxies while blue symbols indicate spirals. Symbol
sizes are proportional to the tidal parameter described in Section
2. Galaxies that contain active AGN are shown using open symbols
and objects that are currently interacting are indicated using
boxes. Small black dots indicate the nearby ($0<z<0.1$), massive
($>L^*$) early-type population drawn from the SDSS following the
method used by K07. Galaxies blueward of $(NUV-r)\sim5.5$
(indicated using the green horizontal line) are very likely to
have had some star formation within the last Gyr (see Section 3 in
K07). BOTTOM: Comparison between the UV colours of the general
early-type population on the optical red sequence ($g-r>0.65$) and
the vD05 sample. See text for more discussion.}
\label{fig:colours}
\end{figure}

The top panel of Figure \ref{fig:colours} shows the UV and optical
colour-magnitude relations for the vD05 objects. Small black dots
indicate the nearby ($0<z<0.1$), massive ($>L^*$) early-type
population, drawn from the SDSS DR4 following the method used by
K07. {\color{black}Briefly, early-type galaxies are selected from
the SDSS using the pipeline parameter \texttt{fracdev}, which
provides a probabilistic measure of how well a `de Vaucouleurs'
profile fits the galaxy's light profile. Galaxies with
$\texttt{fracdev}>0.95$ are extracted and then visually inspected
to remove contaminants such as face-on spirals.} Galaxies blueward
of $(NUV-r)\sim5.5$ (indicated by the green horizontal line) are
very likely to have had some star formation within the last Gyr
(see Section 3 in K07). It is clear that, while the galaxies in
this sample are optically red, they exhibit a wide spread in UV
colours.

{\color{black}We note that a central AGN could contaminate the UV
spectrum of a galaxy, perhaps enhancing the blue UV colour.
However, the contamination from a Type II AGN is likely to be less
than around 15\% in UV flux, which translates to around 0.15 mags
in the $(NUV-r)$ colour, much smaller than the spread in the UV
CMR ({\color{black}$\sim 6$ mags}; see Salim et al. 2007).
Typically, blue early-type colours ($NUV-r<5.5$) are not
restricted to galaxies hosting AGN, either in the sample studied
here or in the general early-type population studied by K07.
Furthermore, the quality of the SED fitting that is used to
calculate star formation histories (see the next section) is
equally good in galaxies which carry BPT signatures of AGN and
those that do not show any signs of AGN, indicating that there is
no measurable contribution from a power-law component. Finally, a
study of the GALEX images of nearby AGN hosts indicates that the
UV emission is extended, making it unlikely that it comes from a
central source (Kauffmann et al. 2007). Thus, the presence of Type
II AGN does not affect the analysis in this paper.}

We begin by comparing the vD05 galaxies to the general early-type
population. For a direct comparison we restrict the analysis to
the K07 optical red sequence ($g-r>0.65$) in the redshift range
$0.05<z<0.1$ and compare the UV colour distribution of the two
samples (bottom panel of Figure \ref{fig:colours}). We find that
the UV distributions are consistent with each other (within
counting errors), although the vD05 sample does show a slight
excess of objects in two intermediate colour bins around
$(NUV-r)\sim4.3$ and $(NUV-r)\sim4.7$.

Recall, however, that the method for selecting the general
early-type galaxy population in K07 involves extracting galaxies
whose light profiles in the $g,r$ and $i$-bands match a `de
Vaucouleurs' profile very closely (by setting
\texttt{fracdev}$>0.95$ when selecting objects from the SDSS
database). To quantify the effect of this criterion on the colour
comparison, we study the variation of the $(NUV-r)$ colour in the
vD05 sample with their \texttt{fracdev} values (inset in the
bottom panel of Figure \ref{fig:colours}). {\color{black}About
57\% of the vD05 galaxies with \texttt{fracdev}$<0.95$ have
$(NUV-r)<5$. The fraction of early-types that were missed from K07
as a result of the \texttt{fracdev}$>0.95$ criterion is around
20\%, from early-type catalogs selected by eye-inspection
\emph{alone} (i.e. no fracdev cut; Kevin Schawinski priv. comm.).
Assuming a red sequence fraction of $\sim70$\% (since K07
estimated that 30\% of the early-type population were blue)
implies an $\sim8$\% ($57\%\times20\%\times70\%$) increase in the
fraction of general early-types with colours blueward of
$NUV-r\sim5$.}

This correction removes most of the discrepancy in the UV colour
distributions. However, the fraction of objects in the colour bin
around $(NUV-r)\sim4.3$ remains inconsistent within one-sigma
uncertainties, {\color{black}although the histograms are
consistent at the two-sigma level}. {\color{black}A simple
Kolmogorov-Smirnov (KS) test \citep[see e.g.][]{Wall1996}
indicates that the probability that the two populations are drawn
from the same parent distribution is $\sim$72\%. A similar
conclusion is drawn if we use a Kuiper test (which maintains
sensitivity near the tails). Thus, we find good overall agreement
between the colour distributions, with the difference plausibly
caused by cosmic variance, since the sky coverage of the SDSS DR4
(from which the early-type population is derived) is more than two
orders of magnitude larger than that of the vD05 sample.}

\begin{figure}
\begin{center}
\includegraphics[width=3.5in]{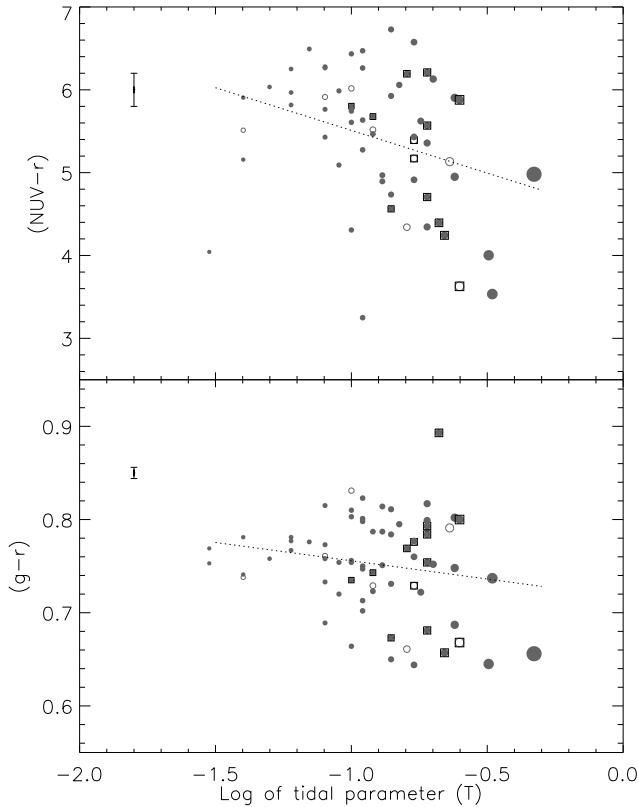}
\caption{Tidal parameter plotted against $(NUV-r)$ and $(g-r)$
colours. {\color{black}A statistically significant correlation
exists between the tidal parameter and both the UV and optical
colours (see text in Section 3.3 for details)}. The symbol size
scales with the size of the tidal distortion. Galaxies that host
Type II AGN are shown using open symbols. Bluer galaxies show
higher tidal distortions, with the trend being much stronger in
the $(NUV-r)$ colour, due to its higher sensitivity to small
amounts of recent star formation.} \label{fig:colour_tidality}
\end{center}
\end{figure}


\subsection{Correlation between colour and tidal distortion}
{\color{black}In Figure \ref{fig:colour_tidality} we plot the UV
and optical colours against the tidal parameter. Note that the
vertical scale in the top panel of this figure, which shows the
$(NUV-r)$ colour, {\color{black}spans 4 mags}, while that for the
$(g-r)$ {\color{black}colour spans only 0.3 mag}. We explore the
strength of the correlation between colour and tidal distortion
using Spearman's rank correlation test, which produces a
non-parametric measure ($\rho$) of the statistical dependence
between two variables. The value of $\rho$ varies between +1 and
-1, which indicate perfect positive and negative correlations
respectively. $\rho\sim0$ indicates a complete lack of correlation
between the variables.

We find that, for the relationship between $(NUV-r)$ and the tidal
parameter, $\rho=-0.43$. {\color{black}The significance of this
value is 0.0032. This is the probability of finding a correlation
between $(NUV-r)$ and the tidal parameter purely through
statistical accident. Since this number is very small, this
implies that the correlation is statistically significant.} The
corresponding values for the $(g-r)$ colour are $\rho=-0.08$, with
a significance of 0.35. {\color{black}Note that the probability of
a chance correlation is higher for the optical colour essentially
due to the correlation being weaker.} We find, therefore, that a
measurable correlation exists between both the optical and UV
colours and tidal distortion in the galaxy. Not unexpectedly, the
correlation is much stronger (and has higher confidence) in the UV
colour, since it is much more sensitive to small amounts of RSF.

The reasonably large scatter in the relationship between colour
and tidal distortion can be understood by noting that the stellar
populations that are responsible for the tidal features and those
that drive the UV colours tend to be different. Assuming that the
RSF is driven by merging (see also Kaviraj et al. 2010), numerical
simulations (e.g. Kaviraj et al. 2009) indicate that the tidal
features form from the underlying stellar populations of the
accreted satellite, while the UV colour is sensitive to the gas
content of the satellite and how far star formation has progressed
when the remnant is viewed. For example, two satellites with
similar underlying stellar populations but different gas contents
might yield similar tidal distortions (if viewed at the same point
during the merger process) but exhibit different UV colours. Thus,
while a high tidal distortion should, in general, be expected to
coincide with a blue UV colour because the remnant is viewed at an
earlier phase in the merger/star formation process and the stellar
populations are younger, the spread in the satellite gas contents
will induce a scatter into the correlation.}

\section{Recent star formation histories}
\subsection{Parameter estimation}
A key objective of this work is to quantify the RSF in the vD05
sample. We are primarily interested in exploring the age of the
last star formation event (which is presumably linked to the
morphological disturbances observed in the deep optical images)
and the stellar mass fractions contributed by these events.

We estimate parameters governing the star formation history (SFH)
of each galaxy by comparing its $(FUV, NUV, u, g, r, i, z)$
photometry to a library of synthetic photometry, generated using a
large collection of model SFHs, {\color{black}specifically
optimized for studying spheroidal galaxies at low redshift. As we
describe below, our scheme decouples the RSF episode from the star
formation that creates the bulk, underlying population. We choose
a parametrisation for the model SFHs that both minimises the
number of free parameters and captures the macroscopic elements of
the star formation history of spheroidal galaxies in the
low-redshift Universe.

Since the underlying stellar mass in spheroidals forms at high
redshift and over short timescales, we model the underlying
stellar population using an instantaneous burst at high redshift.
We put this first (primary) instantaneous burst at z=3. Note that
changing this to z=2, or even z=1, does not affect our conclusions
about the RSF, because the first burst does not contribute to the
UV. A large body of recent evidence suggests that the star
formation in these systems in the \emph{nearby} Universe is driven
by minor mergers \citep[see][and references therein]{Kaviraj2010}.
This star formation is bursty and we model the RSF episode using a
second instantaneous burst, which is allowed to vary in age
between 0.001 Gyrs and the look-back time corresponding to $z=3$
in the rest-frame of the galaxy, and in mass fraction between 0
and 1. Our parametrisation is similar to previous ones used to
study elliptical galaxies at low redshifts
\citep[e.g.][]{Ferreras2000}.} {\color{black}We briefly note that
our scheme differs in construction from other recent work that
explore the general galaxy population at higher redshifts, without
reference to morphology \citep[e.g.][]{Muzzin2009}. In particular,
the decoupling employed in this study between RSF and the bulk
stellar population is difficult to apply at high redshift, where
the underlying stellar populations are actively being assembled.
Thus, our particular parametrisation works best in spheroidal
galaxies at low redshift and is not directly comparable to more
general schemes, especially in the high redshift regime.}

To build the library of synthetic photometry, each model SFH is
combined with a single metallicity in the range 0.1Z$_{\odot}$ to
2.5Z$_{\odot}$ and a value of dust extinction parametrised by
$E_{B-V}$ in the range 0 to 0.5. The dust model employed in this
study is the empirical dust prescription of \citet{Calzetti2000}.
Photometric predictions are generated by combining each model SFH
with the chosen metallicity and $E_{B-V}$ values and convolving
with the stellar models of \citet{Yi2003} through the GALEX and
SDSS filtersets. The model library contains $\sim750,000$
individual models. {\color{black}Note that the $FUV$ and $NUV$
filters have effective wavelengths of $\sim1500 \mathrm{\AA}$ and
$\sim2300\mathrm{\AA}$ respectively.

Since our galaxy sample spans a range of redshifts, equivalent
libraries are constructed at redshift intervals of $\delta
z=0.01$. A fine redshift grid is essential in such a low redshift
study because a small change in redshift produces a relatively
large change in look-back time over which the $UV$ flux can change
substantially, inducing `K-correction-like' errors into the
analysis.

The primary free parameters in this analysis are the age ($t_2$)
and mass fraction ($f_2$) of the second burst (the mass fraction
of the primary burst is simply $1-f_2$). Secondary parameters of
interest are the dust properties of the system. In particular, the
dust content of the primary burst, $E^{PB}_{B-V}$, (which is
effectively tracing the `underlying' population of the galaxy)
should provide a measure of the extinction in the inter-stellar
medium (ISM), while the dust in the secondary burst
($E^{SB}_{B-V}$) should reflect the dust in the star formation
regions and their vicinity.

In each case, the value of the free parameters are estimated by
comparing each observed galaxy to every model in the synthetic
library, with the likelihood of each model ($\exp -\chi^2/2$)
calculated using the value of $\chi^2$, computed in the standard
way. {\color{black}In addition to the observational errors for
each object, we assume additional uncertainties that account for
uncertainties in the models (e.g. Yi 2003), model offsets from
observational data (e.g. Eisenstein et al. 2001; Maraston et al.
2009) and systematic errors in GALEX and SDSS photometry (see e.g.
Ivezi{\'c} et al. 2004; Morrissey et al. 2007). These are taken to
be 0.05 mags for the optical filters, 0.2 mags in $FUV$ and 0.1
mags for the $NUV$ passband. The two types of uncertainty are
added in quadrature.} From the joint probability distribution,
each parameter is marginalised to extract its one-dimensional
probability density function (PDF). We take the median of this PDF
as the best estimate of the parameter in question and the 16 and
84 percentile values as the `one-sigma' uncertainties on this
estimate. In the analysis that follows we present these median
parameter values.

{\color{black} The leverage in $t_2$ and the quality of the $t_2$
fits depends critically on our access to the rest-frame UV, which
hosts most of the flux from hot, young main sequence stars. Note
that \citet{Whitaker2008} have demonstrated that the spheroidal
galaxies in the vD05 sample are virtually dustless, so that there
should be very little reprocessing of the UV-optical light into
the infrared passbands. \citet{Muzzin2009} have shown that the
inclusion of restframe near-infrared (NIR) filters does not alter
the mean values of fitted parameters (although they do reduce the
derived uncertainties) for galaxies in the high-redshift ($z \sim
2$) Universe, which should be significantly dustier than our
low-redshift spheroidal sample. Since our sample is dust-poor, it
seems reasonable, therefore, to suggest that the addition of NIR
filters will leave our conclusions regarding the derived
parameters unchanged. Finally, a similar analysis on a small
sample of SDSS early-type galaxies which have, in addition to
GALEX, NUV photometry from the XMM `Optical Monitor (OM)' (the OM
has three NUV filters around GALEX NUV, see Mason et al. 2001)
indicates that a larger number of UV filters leaves the median
values of $t_2$ and $f_2$ unchanged, with the uncertainties
decreasing by $\sim$10\%. Thus the size of our filterset, both in
the UV and in the optical spectrum, are optimal for the aims of
this paper.}


\subsection{Star formation history parameters}
\subsubsection{Age and mass fraction of the second burst}
Figure \ref{fig:f2_t2} shows the age ($t_2$) and mass fraction
($f_2$) of the RSF estimated to have formed in the second burst in
each of the vD05 early-types. Galaxies are colour-coded based on
their ($NUV-r$) colour and average uncertainties in the parameters
are indicated in the plot. Note that, although the UV colour
changes rapidly with the value of $t_2$, the same colour can be
degenerate with a wide range of mass fractions $f_2$ (see Figure 1
in Kaviraj 2008). Since this effect becomes more pronounced as the
starburst ages, the mass fraction errors become larger at higher
ages. As a result, while we can typically constrain the \emph{age}
of the second burst with good accuracy, the uncertainty in the
mass fraction could span almost a decade.

We find that the bulk of the objects in our sample have
experienced a star formation event in the last 3 Gyrs. Considering
galaxies which have well-constrained mass fraction uncertainties,
the median values of $f_2$ are typically less than 10\%. 5\% of
the early-type sample has experienced a star-formation event
within the last 0.1 Gyrs. The very low mass fractions in this
subset may indicate that the star formation in these events is
still ongoing.

{\color{black}Note that the median mass ratio of the ongoing
mergers in the vD05 sample is 1:4. If the ongoing mergers are
representative of the red merger population, then less than 25\%
of the new stellar mass in the remnants is contributed by star
formation.} The rest is `dry accretion' of existing stars in the
progenitor galaxies. Note, however, that the errors in the mass
fraction estimates are typically quite large and this statement
probably carries with it some uncertainty. Using the lower limit
of the error bars in the mass fraction would suggest a lower
fraction of less than 10\% being contributed by star formation,
{\color{black}which is consistent with the recent literature
\citep[e.g.][]{VD2010}}.

\begin{figure}
\begin{center}
\includegraphics[width=3.5in]{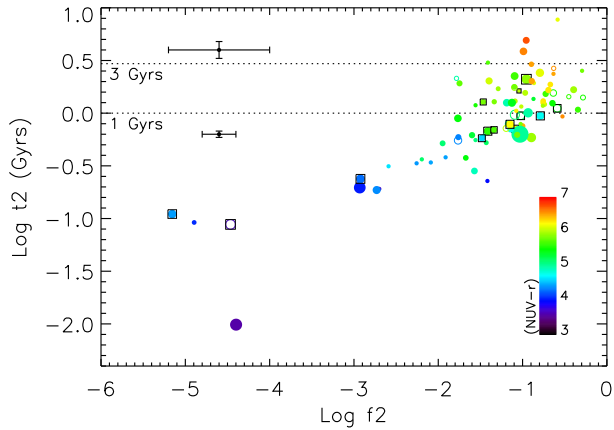}
\caption{Ages ($t_2$) and mass fractions ($f_2$) of the second
star formation episode in the early-type galaxies studied here.
{\color{black}The bulk of the vD05 sample has experienced a star
formation event in the last 3 Gyrs, which typically adds less than
10\% to the stellar mass of the galaxy.} Symbol sizes are
proportional to the tidal parameter described in Section 2.
Galaxies that contain active AGN are shown using open symbols and
objects that are currently interacting are indicated using boxes.
The galaxies are shown colour-coded based on their $(NUV-r)$
colour.} \label{fig:f2_t2}
\end{center}
\end{figure}

{\color{black}In  Figure \ref{fig:t2f2_tidality}, we plot the age
($t_2$) and mass fraction ($f_2$) of the second burst against the
tidal parameter. Recall that $t_2$ is effectively the look-back
time to the last star formation event in the galaxy. We find that
galaxies with larger tidal distortions show lower values of $t_2$,
consistent with the expectation that the intensity of the tidal
distortion decreases with time, resulting in the observed
correlation between the tidal parameter and $t_2$. The best-fit
relation is described by:

\begin{equation}
\log{T} \sim -0.91^{\pm0.03} - 0.21^{\pm0.05}\times \log{t_2}
\hspace{0.05in} {\color{black}\textnormal{[Gyrs]}}
\end{equation}

\begin{figure}
\begin{center}
$\begin{array}{c}
\includegraphics[width=3.5in]{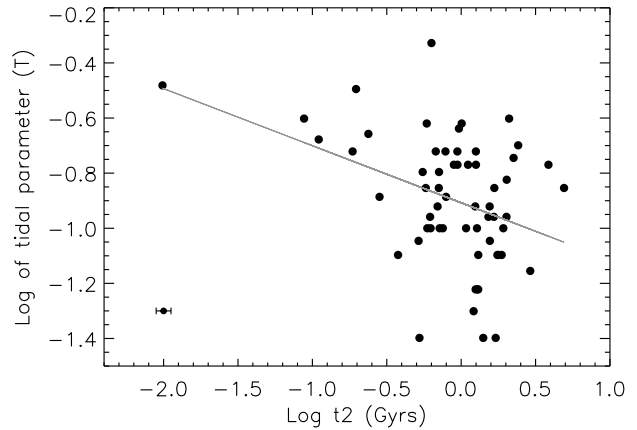}\\
\includegraphics[width=3.5in]{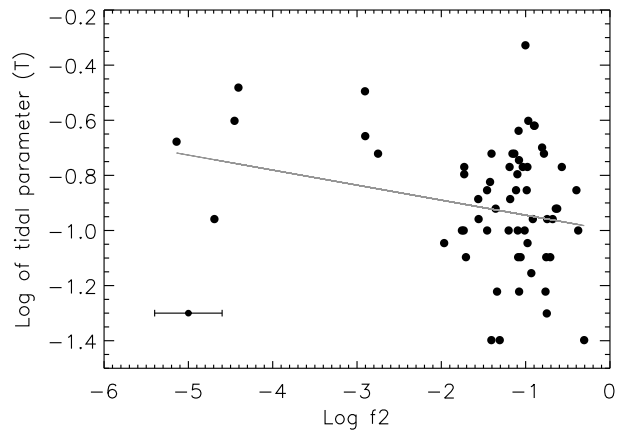}
\end{array}$
\caption{{\color{black}The age ($t_2$) and mass fraction ($f_2$)
of the recent star formation in the vD05 sample, plotted against
the strength of their tidal distortions. Galaxies that have
commenced their star formation more recently (i.e. lower values of
$t_2$) show larger tidal distortions. A weak trend also exists
between the mass fraction $f_2$ and the tidal parameter, in the
sense that smaller mass fractions correspond to higher values of
the tidal parameter.}} \label{fig:t2f2_tidality}
\end{center}
\end{figure}

A weaker correlation is found between the mass fraction formed in
the recent starburst ($f_2$) and the tidal parameter, as shown in
the bottom panel of Figure \ref{fig:t2f2_tidality}. Note that the
range on the x-axis in this plot is much larger than that in the
$t_2$ vs tidal parameter plot (top panel). Since the tidal
features are composed of the the underlying stellar mass of the
accreted satellite and not the young stars emerging from the
starburst (which tend to form preferentially in the central
regions of the remnant), the tidal distortion should be relatively
insensitive to $f_2$. However, a residual correlation should be
expected, since a larger tidal distortion indicates that the
merger is being observed at an earlier stage in the relaxation
process. Hence the mass fraction formed is smaller than what it
would be if the same remnant was observed later, when the tidal
distortion would have decreased and a larger portion of the star
formation would have been completed.

The observed trends between colour, $t_2$, $f_2$ and tidal
distortion can be explained by the fact that galaxies with higher
tidal distortions are at an earlier stage in the merger process
and the star formation episode that is induced by it. As time
passes the remnant relaxes and the star formation activity induced
by the merger decreases, as the gas fuelling this process
gradually runs out. This results both in lower tidal distortions,
redder colours, and increasing mass fractions, accounting for the
trends observed in this study.

{\color{black}We note that our results are in good agreement with
the recent study of S{\'a}nchez-Bl{\'a}zquez et al. (2009), who
studied the vD05 sample using spectral line diagnostics.}


\subsubsection{Metallicity and dust properties}
{\color{black}The median metallicity values for the galaxies in
our sample are in the range 0.8 to 1.4 Z$_{\odot}$, as might be
expected for spheroidal galaxies. Recall, from Section 4.1, that a
single metallicity is fitted to the entire galaxy, so that the
derived metallicity is a represenative value for the combined
stellar population (old + young) in the galaxy.}

{\color{black}In Figure \ref{fig:pbebv_sbebv} we present the dust
properties of the galaxies in our sample. We find that the ISM
extinction in these objects, which is traced by the dust content
of the primary burst ($E^{PB}_{B-V}$), is generally less than 0.1,
consistent with the expected properties of early-type galaxies as
relatively dust-poor systems. \emph{In agreement with the findings
of \citet{Whitaker2008}, we find that the redness of the vD05
objects is not due to large amounts of dust but simply because the
bulk of their stellar populations is old.}}

While the macroscopic ISM seems to be relatively dust-free, we
typically derive larger values for $E^{SB}_{B-V}$ (which is
representative of the dust in star-forming regions in the galaxy).
The bulk of the sample satisfies $E^{SB}_{B-V}<0.4$. Figure
\ref{fig:pbebv_sbebv} indicates that the values of $E^{SB}_{B-V}$
can be a few factors larger than $E^{PB}_{B-V}$, which is
consistent with the fact that the extinction around star forming
regions (specifically in molecular clouds) is expected to be
several times larger than that due to the ISM alone
\citep[e.g.][]{Charlot2000}. The mean value of
$E^{PB}_{B-V}/E^{SB}_{B-V}$ is $\sim3$ (note that the theoretical
estimate, \emph{for starburst galaxies}, derived by
\citet{Charlot2000} is also $\sim3$). While $E^{SB}_{B-V}$ applies
to only a minority ($<10$\%) of the stellar mass in the galaxy, it
is worth noting that the higher extinction appears to affect the
second burst even when $t_2$ is substantially greater than the
ages of molecular clouds, which are typically 10-30 Myrs
\citep[e.g][]{Blitz1980,Hartmann2001}. This may indicate that
regions of the galaxy where the RSF originates remain dusty. In
addition, the parametrisation used in this study does not take
into account the timescale over which the RSF takes place (since
all bursts are assumed to be instantaneous). In reality, only the
youngest stars in the RSF age distribution will probably suffer
high extinction but the lack of a timescale in our analysis leads
to the entire second burst suffering higher extinction to account
for the plausible effect of molecular clouds.

\begin{figure}
\begin{center}
\includegraphics[width=3.5in]{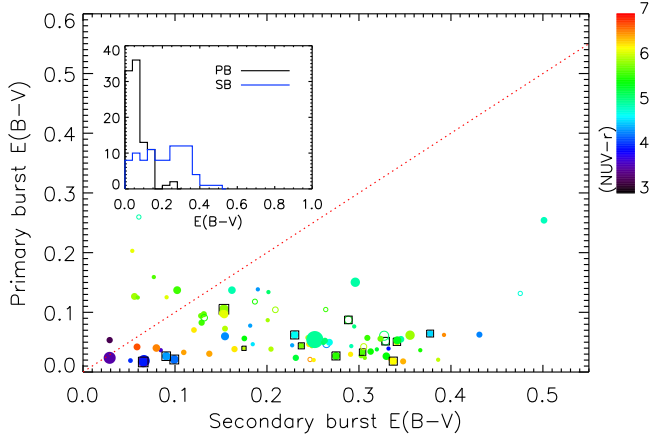}
\caption{The dust content of the primary burst plotted again the
dust content of the secondary burst. {\color{black}The dust
extinction of the primary burst (which represents the ISM) is
typically less than 0.1, consistent with early-types being
dust-poor. The extinction in star-forming regions is, on average,
a factor of 3 higher.} Symbol sizes are proportional to the
tidality parameter described in Section 2. Galaxies that contain
active AGN are shown using open symbols and objects that are
currently interacting are indicated using boxes. The galaxies are
colour-coded based on their ($NUV-r$) colour.}
\label{fig:pbebv_sbebv}
\end{center}
\end{figure}


\subsection{A prescription to estimate the age of the last star
formation event} The parameter estimation performed in this study
allows us to develop a recipe to estimate the age of the most
recent star formation event, given the UV and optical photometry
of the galaxy in question. We focus on a recipe of the age rather
than the mass fraction because the smaller uncertainties in the
age make the age estimation significantly more robust.

We note first that the median value for $t_2$ used in this
analysis is extracted from its marginalised PDF, so that the
effects of all other parameters (mass fraction, dust and
metallicity) have been integrated out. In the top panel of Figure
\ref{fig:t2_prescription}, we plot the median value of $t_2$
against the $(NUV-r)$ colour. We also show the equivalent plot for
the optical $(g-r)$ colour in the inset.

While a clear correlation exists between $t_2$ and the $(NUV-r)$
colour, there is no equivalent trend with the optical colour. This
plot demonstrates both the usefulness of the UV spectrum in
quantifying the age of the RSF in early-type galaxies and the
insensitivity of the optical spectrum to the residual star
formation in these objects.

\begin{figure} $\begin{array}{c}
\includegraphics[width=3.5in]{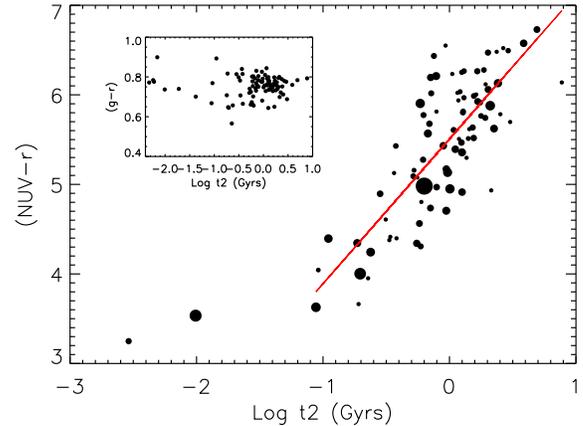}\\
\includegraphics[width=3.5in]{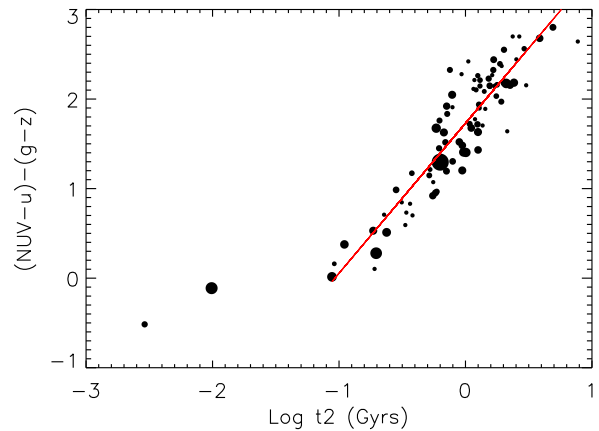}
\end{array}$
\caption{TOP: The ($NUV-r$) colour plotted against the age of the
second burst. BOTTOM: The relatively `dust-insensitive'
$(NUV-u)-(g-z)$ double colour plotted against the age of the
second burst. The solid line indicates a linear least-squares fit
to the relation, {\color{black}which allows us to estimate the age
of the last star formation event in early-type galaxies directly
from their observed $NUV$, $u$, $g$ and $z$-band photometry.}}
\label{fig:t2_prescription}
\end{figure}

Although a correlation exists between $t_2$ and ($NUV-r$), we note
that the ($NUV-r$) colour is dust sensitive. Standard dust
extinction laws \citep[e.g.][]{Calzetti2000,Cardelli1989} indicate
that the extinction in the ($NUV-r$) colour, $A^{NUV-r}\sim 6
\times E_{B-V}$. Recalling that the estimate for $t_2$ is
marginalised over all other parameters (including dust), we would,
ideally, like to compare it to an observational quantity which is
also insensitive to dust.

Since the extinctions in colours are linear combinations of the
extinctions in the constituent filters, we construct a double
colour, $(NUV-u)-(g-z)$, that minimises the dust sensitivity,
given the filter combinations available to us for the bulk of the
vD05 sample. The extinction in this colour (assuming a
\citet{Calzetti2000} extinction law) is $A^{(NUV-u)-(g-z)}\sim 1.5
\times E_{B-V}$, a factor of 4 less than that in $(NUV-r)$. Not
unexpectedly, plotting $t_2$ against $(NUV-u)-(g-z)$ produces the
same quantitative trend but with less scatter (bottom panel of
Figure \ref{fig:t2_prescription}). The best-fit line (shown using
the solid red line) yields:

\begin{equation}
\log t_2 \hspace{0.05in} {\color{black}\textnormal{[Gyrs]}} \sim
0.6^{\pm0.03}\times[(NUV-u)-(g-z)-1.73^{\pm0.03}]
\end{equation}

Eqn (2) allows us to estimate the age of the last star formation
event in early-type galaxies, directly from their \emph{observed}
(i.e. not K-corrected) $NUV, u, g$ and $z$-band photometry. We
note that there appears to be a possible flattening of the slope
in the correlation for the very bluest ($NUV-r<3.8$ or
$(NUV-u)-(g-z)<0$) objects {\color{black}(which have been
{\color{black}excluded} from the fitting above)}. These objects
may obey a different scaling, although the small number of
galaxies in this part of the parameter space makes it difficult to
derive a robust conclusion. It is possible that, given the very
small values of $t_2$, the star formation activity in these events
is in its very early stages, displacing them from the main locus.


\section{Conclusions}
We have combined GALEX (UV) and SDSS (optical) photometry to
derive the recent star formation histories of a sample of
$\sim100$ red sequence galaxies, many of which exhibit widespread
signs of disturbed morphologies (in deep optical imaging) that are
consistent with recent merging events. Originally studied by van
Dokkum (2005; vD05 hereafter), more than 70\% of bulge-dominated
galaxies in this sample show tidal features at a surface
brightness limit of $\mu\sim28$ mag arcsec$^{-2}$. An analysis of
the spatial (RA, DEC and redshift) distributions of the vD05
objects indicates that the sample is not drawn from overdense
regions such as cluster cores and can be considered to be part of
the general `field'. While the vD05 galaxies are optically red,
they show a wide dispersion in UV colours that is attributable to
low level RSF, akin to what has been found recently in the general
early-type population. {\color{black}Comparison of the vD05 UV
colour distribution to that of the general early-type red sequence
(drawn from the SDSS) using a KS test indicates that they are
drawn from the same parent distribution.}

{\color{black}A statistically significant correlation is found
between galaxy colour and the intensity of the tidal distortion,
in the sense that redder objects show smaller tidal distortions. A
stronger trend is observed in the UV colour, since it is more
sensitive to low-level recent star formation. Not unexpectedly, a
correlation also exists between the parametrised look-back time to
the recent star formation event ($t_2$) and the tidal distortion,
in the sense that the tidal distortion decreases with increasing
values of $t_2$. Only a weak trend exists between the mass
fraction formed in the event ($f_2$) and the tidal parameter, with
larger values of the tidal parameter corresponding to smaller
values of $f_2$. The observed trends can be explained by the fact
that galaxies with higher tidal distortions are at an earlier
stage in the merger process and the star formation episode that is
induced by it. As time passes the remnant relaxes and the star
formation activity decreases, as the gas fuelling this process
gradually runs out. This results in lower tidal distortions,
redder colours and increasing mass fractions, as found in the
trends observed in this study.}

The bulk of the vD05 sample has experienced a star formation event
in the last 3 Gyrs. The median values of the mass fractions are
typically less than 10\%. 5\% of the sample has experienced a
star-formation event within the last 0.1 Gyrs and the star
formation episodes in these objects appear to be ongoing since the
mass fractions formed in these events is very small ($<0.1$\%). If
the ongoing mergers in the vD05 sample, that appear to have a
median mass ratio of 1:4 (see Section 6.2 in vD05), are
representative of nearby red mergers, then less than $\sim$25\% of
the new stellar mass in the remnants is contributed by star
formation. The rest is `dry accretion' of existing stars. We
should note, however, that the large uncertainties on the mass
fraction estimates (see Figure \ref{fig:f2_t2}) makes this
statement somewhat uncertain and using the lower limit of the
error bars on the $f_2$ values suggests a lower fraction of less
than 10\% being contributed by star formation.

The parameter estimation performed in this study allows the
primary burst (which takes place at $z=3$ and contributes the
underlying stellar population) and the secondary burst (which
represents the most recent burst of star formation and dominates
the UV flux) to have different dust extinction values. We find
that the dust extinctions in the primary burst ($E^{PB}_{B-V}$)
are typically less than 0.1, consistent with early-type galaxies
having ISMs that are dust-poor. However, while the macroscopic ISM
seems to be relatively dust-free, we typically derive larger
values for the dust extinction in the second burst
($E^{SB}_{B-V}$), with the bulk of the sample satisfying
$E^{SB}_{B-V}<0.4$. The mean value of $E^{SB}_{B-V}/E^{PB}_{B-V}$
is $\sim3$. We note, however, that, while $E^{SB}_{B-V}$ applies
to only a minority ($<10$\%) of the stellar mass in the galaxy,
the higher extinction appears to affect the second burst even when
its age ($t_2$) is substantially greater than the ages of
molecular clouds, which are typically 10-30 Myrs. This may
indicate that regions of the galaxy where the RSF originates
remains dusty. It could also be a result of the parametrisation
used here, which does not take into account the timescale over
which the RSF takes place (since all bursts are assumed to be
instantaneous). In reality, only the youngest stars in the RSF age
distribution will probably suffer high extinction but the lack of
a timescale in our analysis leads to the entire second burst
suffering higher extinction to account for the plausible effect of
molecular clouds.

We have used our analysis to develop a recipe to estimate the age
of the most recent star formation event, given the \emph{observed}
UV and optical photometry of the galaxy in question. We find a
well-defined correlation between the age of the second burst
($t_2$), which is marginalised over all other free parameters and
the relatively dust-insensitive double colour, $(NUV-u)-(g-z)$.
Eqn (2) allows us to estimate the look-back time to the last star
formation event in red early-type galaxies, \emph{directly} from
their \emph{observed} (i.e. not K-corrected) $NUV, u, g$ and
$z$-band photometry. The slope in the correlation appears to
flatten for the very bluest ($NUV-r<3.8$ or $(NUV-u)-(g-z)<0$)
objects, possibly due to the RSF being ongoing in these systems.

{\color{black}Since the vD05 sample appears to be representative
of the general field population of spheroidal galaxies in the
nearby Universe (Section 3.2), the results of this study suggest
that the widespread low-level star formation that has been
reported in nearby spheroidals is driven by recent mergers. The
vast majority of these events do not involve large amounts of gas,
indicating that they are either dry major mergers or minor
mergers, where the spheroid accretes a small gas-rich satellite
(see also Kaviraj et al. 2010). Nevertheless, the spheroidal
galaxy population, at present-day, is constantly evolving, both
dynamically through interaction with companions and in terms of
low-level star formation activity induced by these interactions.}


\nocite{Martin2005} \nocite{York2000} \nocite{SDSSDR4}
\nocite{Baldwin1981} \nocite{Kauffmann2003}
\nocite{Schawinski2007a} \nocite{Kaviraj2007a}
\nocite{Kaviraj2007b} \nocite{Kaviraj2008a} \nocite{Kaviraj2008b}
\nocite{Gawiser2006} \nocite{Kauffmann2007} \nocite{Bernardi2003a}
\nocite{Salim2007}


\section*{Acknowledgements}
I am grateful to the referee for a comprehensive and insightful
report which helped strengthen several parts of the paper. I
warmly thank Pieter van Dokkum for numerous discussions across
several iterations of this paper.

I acknowledge a Research Fellowship from the Royal Commission for
the Exhibition of 1851, an Imperial College Research Fellowship
and a Senior Research Fellowship from Worcester College, Oxford.
Part of this work was supported by a Leverhulme Early-Career
Fellowship. Richard Ellis, Sukyoung Yi, Kevin Schawinski and
Vandana Desai are thanked for constructive comments.

GALEX (Galaxy Evolution Explorer) is a NASA Small Explorer,
launched in April 2003, developed in cooperation with the Centre
National d'Etudes Spatiales of France and the Korean Ministry of
Science and Technology.

Funding for the SDSS and SDSS-II has been provided by the Alfred
P. Sloan Foundation, the Participating Institutions, the National
Science Foundation, the U.S. Department of Energy, the National
Aeronautics and Space Administration, the Japanese Monbukagakusho,
the Max Planck Society, and the Higher Education Funding Council
for England. The SDSS Web Site is http://www.sdss.org/.

The SDSS is managed by the Astrophysical Research Consortium for
the Participating Institutions. The Participating Institutions are
the American Museum of Natural History, Astrophysical Institute
Potsdam, University of Basel, University of Cambridge, Case
Western Reserve University, University of Chicago, Drexel
University, Fermilab, the Institute for Advanced Study, the Japan
Participation Group, Johns Hopkins University, the Joint Institute
for Nuclear Astrophysics, the Kavli Institute for Particle
Astrophysics and Cosmology, the Korean Scientist Group, the
Chinese Academy of Sciences (LAMOST), Los Alamos National
Laboratory, the Max-Planck-Institute for Astronomy (MPIA), the
Max-Planck-Institute for Astrophysics (MPA), New Mexico State
University, Ohio State University, University of Pittsburgh,
University of Portsmouth, Princeton University, the United States
Naval Observatory, and the University of Washington.


\nocite{Kaviraj2007c} \nocite{Mason2001} \nocite{Sanchez2009}
\nocite{Eisenstein2001} \nocite{Yi2003} \nocite{Maraston2009}
\nocite{Ivezic2004} \nocite{Morrissey2007}


\bibliographystyle{mn2e}
\bibliography{references}


\end{document}